\documentclass[fleqn,usenatbib]{mnras}

\usepackage{newtxtext,newtxmath}

\usepackage[T1]{fontenc}
\usepackage{ae,aecompl}

\usepackage{graphicx}	
\usepackage{amsmath}	
\usepackage{amssymb}	
\usepackage{subfigure}

\title[Dynamics of M31 Dwarf Galaxies]{On the origin of the asymmetric dwarf galaxy distribution around Andromeda}

\author[Z Wan et al.]{Zhen Wan$^{1}$\thanks{E-mail: zhen.wan@sydney.edu.au},
William H. Oliver$^{1}$,
Geraint F. Lewis$^{1}$,
Justin I. Read$^{2}$ and
\newauthor
Michelle L. M. Collins$^{2}$
\\
$^{1}$Sydney Institute for Astronomy, School of Physics A28, The University of Sydney, NSW, 2006, Australia\\
$^{2}$Department of Physics, University of Surrey, Guildford, GU2 7XH, Surrey, UK
}

\date{Accepted XXX. Received YYY; in original form ZZZ}

\pubyear{2019}

\usepackage{etoolbox}
\makeatletter
\patchcmd\@combinedblfloats{\box\@outputbox}{\unvbox\@outputbox}{}{%
   \errmessage{\noexpand\@combinedblfloats could not be patched}%
}%
 \makeatother

\begin{document}
\label{firstpage}
\pagerange{\pageref{firstpage}--\pageref{lastpage}}
\maketitle

\begin{abstract}
The dwarf galaxy distribution surrounding M31 is significantly anisotropic in nature. Of the 30 dwarf galaxies in this distribution, 15 form a disc-like structure and 23 are contained within the hemisphere facing the Milky Way. Using a realistic local potential, we analyse the conditions required to produce and maintain these asymmetries. We find that some dwarf galaxies are required to have highly eccentric orbits in order to preserve the presence of the hemispherical asymmetry with an appropriately large radial dispersion. Under the assumption that the dwarf galaxies originate from a single association or accretion event, we find that the initial size and specific energy of that association must both be relatively large in order to produce the observed hemispherical asymmetry. 
However if the association was large in physical size, the very high-energy required would enable several dwarf galaxies to escape from the M31 and be captured by the Milky Way. Furthermore, we find that associations that result in this structure have total specific energies concentrated around $E = V_{esc}^{2} - V_{init}^{2} \sim 200^2$ -- $300^2\ \rm{km^2\ s^{-2}}$, implying that the initial velocity and initial position needed to produce the structure are strongly correlated. The overlap of initial conditions required to produce the radial dispersion, angular dispersion, and the planar structure is small and suggests that either they did not originate from a single accretion event, or that these asymmetric structures are short-lived.
\end{abstract}




\section{Introduction}
Early evidence of structures in the distributions of dwarf galaxies dates back several decades when \citet{Lynden-Bell1976} discovered that several globular clusters and dwarf galaxies surrounding the Milky Way (MW) lay in streams of high velocity clouds that were thought to form a planar structure \citep{Lynden-Bell1995}. The suspicion that a great disc of MW dwarf galaxies existed was independently confirmed through follow-up studies \citep{Kroupa2005,Metz2007,Metz2008,Pawlowski2012}. This then raised the question of the origin of such structures since the likelihood that they would assemble from a previously isotropic distribution is extremely small. However, studies have suggested that the structure may not rotate coherently \citep{Cautun2015a, Phillips2015} and also that the statistical relevance of disc configurations is heavily influenced by detection bias \citep{Cautun2015b, Buck2016, Maji2017}. More recent papers have created further tension with these findings claiming that the MW satellites as a whole do not lie in a thin plane, although there is strong evidence that their distribution is anisotropic \citep{GaiaCollaboration2018, Simon2018}.

Such an anisotropic distribution of dwarf galaxies has been seen in M31, with the Pan-Andromeda Archaeological Survey \citep{McConnachie2009}, claiming that 15 of the 27 observed dwarf galaxies  constitute a great disc, all with same sense of rotation about M31 \citep{Ibata2013,Conn2013}. The size of this disc is at least $400\ \rm{kpc}$ in diameter with perpendicular scatter of less than $\sim14\ \rm{kpc}$. Adding to the complexity, \citet{Conn2012,Conn2013} have found that the dwarf galaxies surrounding M31 possess a significant hemispherical anisotropy, with 21 of the 27 dwarf galaxies are contained within the same hemisphere. However, the radial distribution of those dwarf galaxies is less special than the directional distribution, with the distances of the dwarf galaxies to the M31 ranging from $40\ \mathrm{kpc}$ to $400\ \mathrm{kpc}$.

The origin of these anisotropic structures has been the subject of much debate. The nature of the cosmic web imposes some coherence on the accretion of neighbouring galactic structures \citep{Zentner2005,Libeskind2011,Libeskind2015}, and consequently, galaxies often fall into larger structures as part of a group \citep{Read2008,D'Onghia2008,Li2009}. In principal, these effects could be responsible for manifesting dwarf galaxy disc structures that surround larger host galaxies within the $\Lambda$CDM model of cosmology \citep[e.g.][]{Lovell2011,Wang2013,Goerdt2013,Bahl2014,Buck2015,Gillet2015}. However, some studies have suggested that the discrepancy between the anisotropy in observation and simulation is significant and is not easily explained with $\Lambda$CDM cosmologies \citep{Kroupa2005,Ibata2014,Pawlowski2012,Pawlowski2013,Forero-Romero2018,Pawlowski2019}. Other studies advocate that 10 \citep{Cautun2015b}, or even 20 per cent \citep{Shao2016} of $\Lambda$CDM haloes have even more prominent planes than those present in the Local Group. It has been proposed that pairs of large galaxies similar to those in the Local Group can impose a shape alignment of satellite galaxies \citep{Wang2019}. Furthermore, \citet{Libeskind2016} and \citet{Gong2019} have suggested that it is statistically likely for satellite distributions surrounding pairs of galaxies -- such as the M31-MW system -- to be lopsided, though these satellites are primarily on their first infall. Inferring the whereabouts of dwarf galaxies in the past can be challenging, particularly if a dwarf's position is to be tracked over more than a full orbit of its host, due to many dwarfs falling in as part of associations \citep{Lux2010}.

Other attempts to explain disc-like structures of dwarf galaxies from a smaller scale perspective have also been made. \citet{Pasetto2009} have tested the feasibility of disc structures in the Local Group as a result of tidal effects, finding that these could account for the planar structure excluding those tightly bound dwarf galaxies. \citet{Bowden2013} show that in a tri-axial Navarro--Frenk--White (NFW) \citep{Navarro1996} potential it is possible for a thin disc structure to persist over cosmological time-scales if and only if it lies in the planes perpendicular to the long or short axis of a tri-axial halo, else it will double in thickness within $\sim 5\ \rm{Gyr}$. Later \citet{Bowden2014} calculated the life-times of inward falling associations in various potentials and found that asymmetric structures could survive longer than the current age of the universe in the outer regions of nearly spherical potentials.

Given this groundwork, our investigation aims to numerically investigate how a more realistic dynamic potential configuration contributes to the formation of these asymmetric structures. We construct the M31--MW system potential by considering the disc, bulge, and halo components of these galaxies in Sec.~\ref{sec:method}. Then in Sec.~\ref{sec:backwards} we place all observed dwarf galaxies surrounding M31 into this potential at their current positions and integrate backwards with various tangential velocities so as to obtain the orbital properties of each dwarf galaxy. In Sec.~\ref{sec:forward} we also integrate the orbits of numerous dwarf galaxy associations forward in time to identify the set of initial conditions required to assemble the currently observed structures surrounding M31 from a single association of dwarf galaxies. Finally, we discuss our findings in Sec.~\ref{sec:discussion}.

\section{Method}
\label{sec:method}
\subsection{Potential} 
To appropriately consider the dynamic behaviour of the dwarf galaxies surrounding M31, a superposition of both the M31 and the MW gravitational potentials are modelled. For the MW, we use the \textsc{MWPotential 2014} in {\it galpy} \citep{Bovy2015}, which is composed with following Eqs. \ref{PowerLaw}, \ref{MiyamotoNagai}, and \ref{NFW}:

We use a spherically symmetric power-law potential with an exponential cut-off for the bulge. This is derived from the mass-density model,
    \begin{equation} \label{PowerLaw}
        \rho(r) = \rho_0r^{-\alpha}\exp\left(\left(-r/r_c\right)^2\right),
    \end{equation}
for which we use a power-law index of $\alpha = 1.8$, and a cut-off radius of $r_c = 1.9\ \rm{kpc}$.

The disc is modelled using the axisymmetric Miyamoto--Nagai potential,
    \begin{equation} \label{MiyamotoNagai}
        \Phi(R,z) = -\frac{\Phi_0}{\sqrt{R^2 + \left(a + \sqrt{z^2 + b^2}\right)^2}},
    \end{equation}
where $R = \sqrt{x^2 + y^2}$ in galacto-centric coordinates. Here we use potential parameters $a = 3\ \rm{kpc}$ and $b = 0.28\ \rm{kpc}$. 

To model the affect of the dark matter halo we use the NFW potential with the density profile,
     \begin{equation} \label{NFW}
        \rho(r) = \frac{\rho_0}{(r/h)(1 + r/h)^2},
    \end{equation}
and characteristic radius $h = 16\ \rm{kpc}$.

This potential is scaled so that the circular velocity at $r = 8\ \rm{kpc}$ away from the galactic centre in the disc ($z = 0\ \rm{kpc}$) is set to $220\ \rm{km\ s^{-1}}$. In addition, the potential parameters are also tuned to match multiple data sets that include observations of the velocity dispersion, vertical force, terminal-velocity, mid-plane density profile slope, and total mass \citep{Dehnen1998,Binney2008,Zhang2013, Bovy2013,Clemens1985,McClure-Griffiths2007,Holmberg2000,Bovy2012,Xue2008}. The corresponding virial mass of this potential is $0.8\times10^{12}\ \mathrm{M_{\odot}}$ which agrees well with the virial mass from dynamical analyses \citep[e.g.][]{Xue2008,Kafle2012,Deason2012,Kafle2014}, and at the low end of a massive Milky Way \citep[e.g.][]{Li2008,Watkins2010, Sohn2018, Posti2019}.

For M31, we use the same potential form as for the MW, however we set $a = 5.09\ \rm{kpc}$, $b = 0.28\ \rm{kpc}$ in Eq.\ref{PowerLaw}, and $h = 20\ \rm{kpc}$ in Eq.\ref{NFW} \citep{Seigar2008}, and the total mass of M31 is 1.5 times of the MW total mass. In the integration, we also examine the effects of the oblate and prolate NFW profiles using the \textsc{TriaxialNFWPotential} with the equi-density radius defined as $r =\sqrt{x^2 + y^2 + (0.5z)^2}$ and $r =\sqrt{x^2 + y^2 + (1.5z)^2}$ respectively for the M31 potential. This increases the asymmetricity of the potential \citep[e.g.][]{Debattista2008,Dubinski1994} and might lead to the anisotropic distribution of the dwarf galaxies \citep{Hayashi2014}. To focus on the shape, we set all the dark halo profiles to have same mass. The mass of the M31--MW system ($\sim 2\times 10^{12}\ \mathrm{M_{\odot}}$) we have constructed is consistent with recent timing argument constraints \citep{Penarrubia2015}. The MW halo is assumed to be spherically symmetric since the shape of the MW potential should be less significant due to the distance between the MW and M31. Furthermore, within $\sim 50\ \rm{kpc}$, the MW halo appears to be rather round \citep[e.g.][]{Ibata1998,Read2014,Wegg2019}.

\begin{figure*}
    \subfigure{\includegraphics[width=\columnwidth]{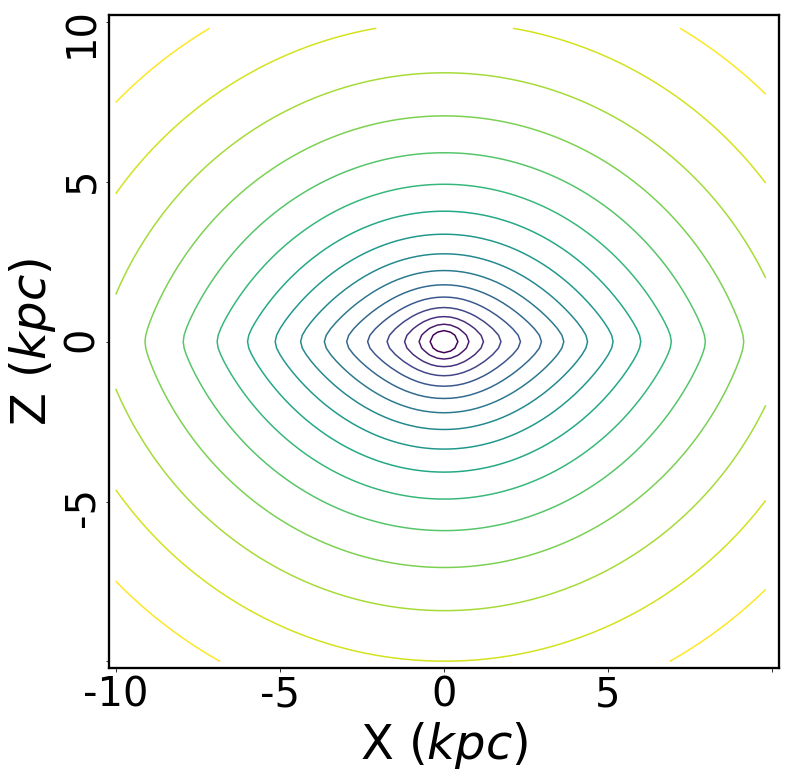}}
    \subfigure{\includegraphics[width=\columnwidth]{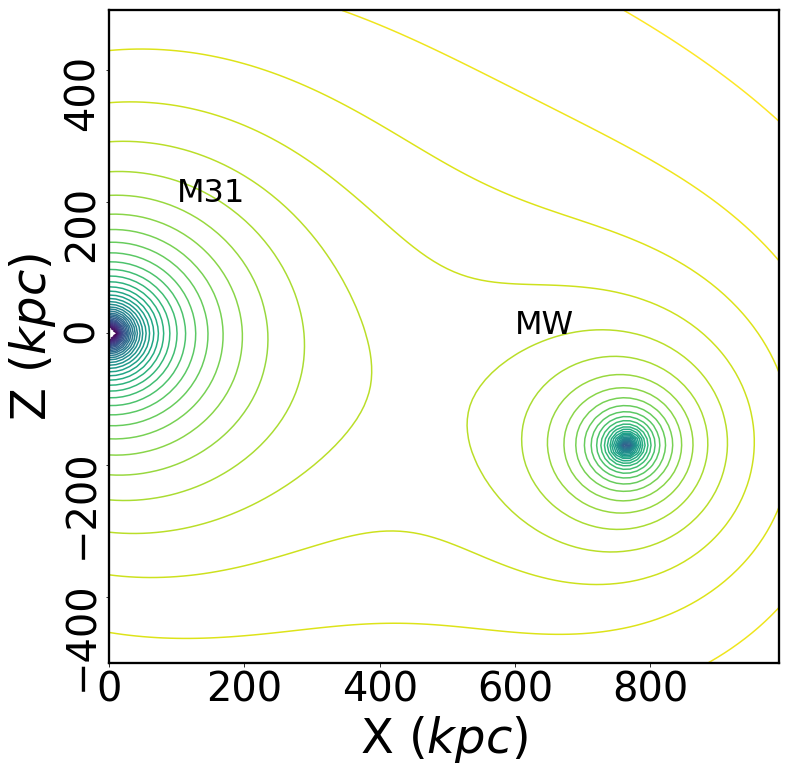}}
        \caption{{\it Left}: Contour lines of equi-potential in X-Z plane. This potential includes a bulge, a disc and a halo, where the bulge and halo are spherical symmetric and the disc is axisymmetric. Here the X-Y plane lies in the disc plane.
        {\it Right}: The potential of the M31--MW system in the X-Z plane, which is centred on the galactic centre of M31. Here the X-Y plane coincides with the M31 disc plane. The MW lies to the right of the figure and due to its presence, a slight deformation of the M31 potential is visible at $\sim 400\ \rm{kpc}$.(see the Fig.~\ref{fig:potential_Triaxial} for the demonstration of the prolate/oblate profiles.)}
        \label{fig:potential}
\end{figure*}

The left panel of Fig.~\ref{fig:potential} depicts the galactic potential contour in X-Z plane within $10\ \rm{kpc}$, of which M31 is at the centre. Then by adopting the M31 configuration as position angle $\theta = 39.8^{\circ}$ and inclination $i = 77.5^{\circ}$ \citep{deVaucouleurs1958,McConnachie2006}, we place another galactic potential at the position of the MW as seen from M31 with the corresponding configuration to include the effect of the MW. The right panel of Fig.~\ref{fig:potential} portrays an overview of the M31--MW system potential we use in our calculations. Here it is clear that the contours deform at $\sim 400\ \rm{kpc}$ from each galaxy's centre.

\subsection{Integration} To calculate the orbits of the dwarf galaxies within the potential we build, we use the \textsc{solve\_ivp} function from the \textit{scipy} package with an adaptive step-size \citep{Eric2001} to solve the differential equation of motion numerically. Here we set the relative tolerance to be $10^{-11}$. Prior to integration, we calculate the current MW velocity based on the M31 line-of-sight velocity from \citet{McConnachie2012} and the solar reflex motion as $(11.1, 232.24, 7.25)\ \rm{km\ s^{-1}}$ \citep{Schonrich2010, Bovy2015}. We then use this relative motion to calculate the position of the MW for all times during the past $10\ \rm{Gyr}$. This then realistically ensures that the effect of the potential from the MW will change dynamically throughout any integration we perform. Given the position and velocity of a dwarf galaxy, we are then able to integrate its orbit both forwards and backwards within the M31--MW potential. In this paper, we ignore the interaction between dwarf galaxies so that the orbits of the dwarf galaxies only depend on the system potential. The results then reflect how the M31--MW potential contributes to the observed anisotropic distribution of the dwarf galaxies surrounding M31.

\section{Results}
\subsection{Backward Integration of Orbits}
\label{sec:backwards}

\begin{figure*}
    \vspace{-0.5cm}
    \includegraphics[trim={28mm 0 28mm 0}, clip, width=1.9\columnwidth]{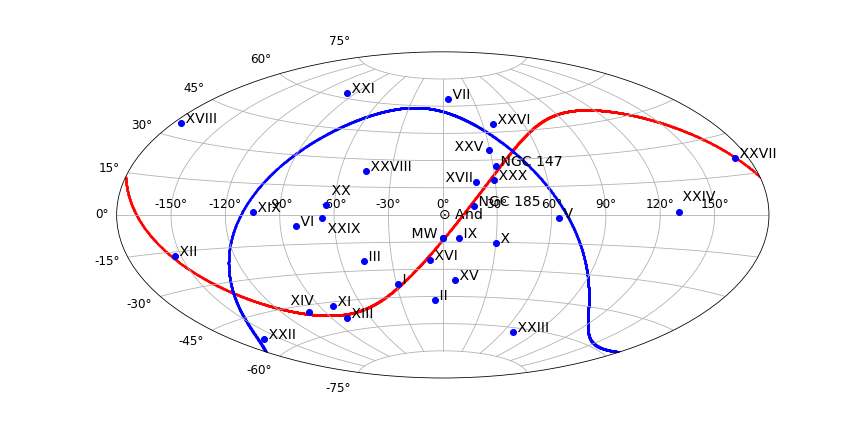}
    \vspace{-0.5cm}
    \caption{An Aitoff projection of the positions of the dwarf galaxies in the M31-centred coordinate system. The positions are taken from \citet{McConnachie2012} and \citet{Conn2012}. The red line indicates the great plane of 15 dwarf galaxies \citep{Ibata2013}, and the blue line separates the hemispheres of largest asymmetry where one side includes the 21 dwarf galaxies in \citet{Conn2013}. We also include a Cartesian projection of the dwarf galaxies in Fig.~\ref{fig:3Dprojection}.}
    \label{fig:aitoff}
\end{figure*}

The position and line-of-sight velocities of the dwarf galaxies surrounding M31 are adopted from \citet{McConnachie2012}, and the distances to each dwarf galaxy are taken from \citet{Conn2012}. We consider these parameters as a current snapshot. Fig.~\ref{fig:aitoff} displays the Aitoff projection of these dwarf galaxies in the M31-centred coordinate system. This figure also indicates the angular asymmetry \citep{Conn2013} as well as the disc structure \citep{Conn2013,Ibata2013}.

This data set however does not provide the tangential components of the velocities of these dwarf galaxies. Without knowledge of these proper motions, we simulate a range of current conditions for each dwarf galaxy. We sample the magnitude of the tangential velocities in $30\ \rm{km\ s^{-1}}$ steps from the interval $30$--$240\ \rm{km\ s^{-1}}$. For each magnitude, the angular direction is also sampled at a resolution of $0.02\ \rm{rad}$ over a full $2\pi$ range. This step size ensures a high resolution as well as an affordable calculation time. In total, 2520 current tangential velocities are sampled for each dwarf galaxy. For each current condition, the dwarf galaxy's orbit is integrated into the past for $10\ \rm{Gyr}$. Note that during this integration, the MW will move away from M31. 

As a measure of orbital frequency, we count the number of times that each dwarf galaxy passes through a pericentre along each orbit integration. Since the current 3-dimensional position and line-of-sight velocity of each dwarf galaxy are fixed, an individual dwarf galaxy's pericentre number is only dependent on the current tangential velocity we assign. The pericentre number tends to be larger for a dwarf galaxy that is currently closer to the centre of M31. It is these same dwarf galaxies whose pericentre number for a particular orbit is more strongly affected by the choice of that orbit's current tangential velocity. Dwarf galaxies closer to the centre of M31 will therefore show a larger range of possible pericentre numbers when compared with dwarf galaxies currently far from the centre of M31. Furthermore, any particular dwarf galaxy's pericentre number will decrease with increasing tangential velocity magnitude. That is until the total velocity reaches the upper limit for bounded orbits of that dwarf galaxy (above which the dwarf galaxy will never pass through a pericentre).
 
\begin{figure*}
    \includegraphics[width=2\columnwidth]{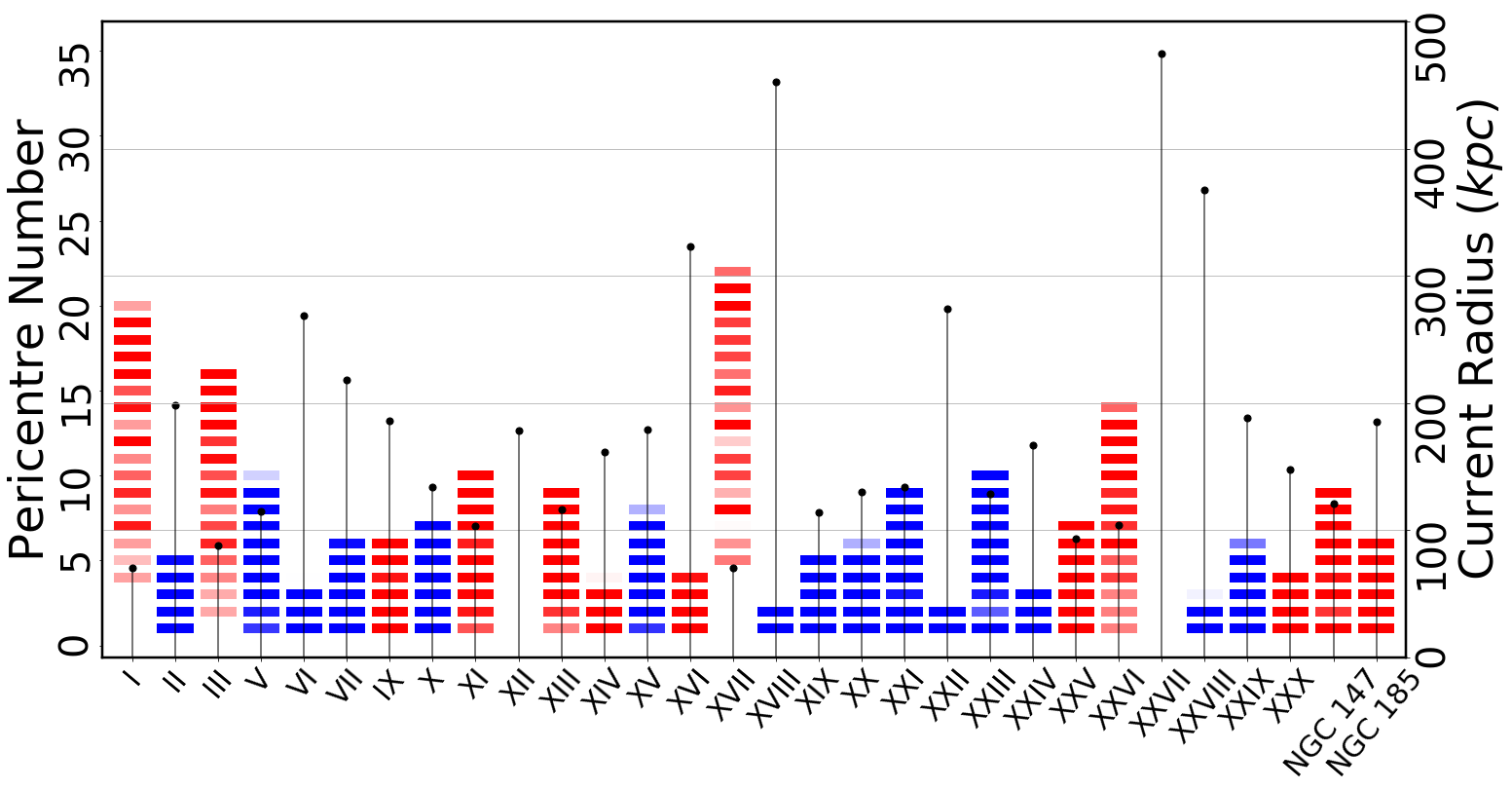}
    \caption{The number of times each dwarf galaxy passes through a pericentre during $10\ \rm{Gyr}$ of backwards integration with different tangential velocities. Here the red bars indicate those 15 dwarf galaxies that belong to the great plane structure (Andromeda XII is considered to be a part of this structure as well). The point markers represent the current radius from M31 of each dwarf galaxy (right axis). It is possible that some dwarf galaxies that are close to the centre of M31 with low line-of-sight velocity could be in low-energy orbit. These dwarf galaxies are closely bound to M31 and will conclude more than 20 passes of M31 during $10\ \rm{Gyr}$. Other dwarf galaxies that are far from M31 may only exist on high-energy orbits and consequentially only be able to pass M31 up to a few times.}
    \label{fig:period}
\end{figure*}

\begin{figure}
    \includegraphics[width=\linewidth]{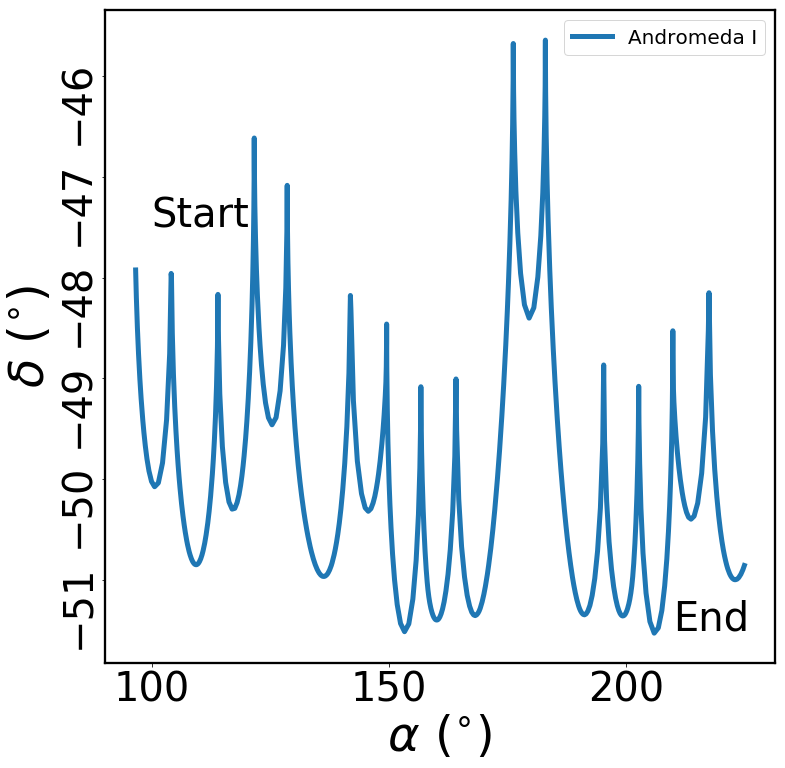}
    \vspace{-0.5cm}
    \caption{The direction of angular momentum of one possible low-energy orbit of Andromeda I over $10\ \rm{Gyr}$ backwards integration. The orientation of the orbit plane changes substantially throughout the integration due to the precession of the angular momentum. This implies that a dwarf galaxy in this orbit is not able to stay within one plane during a $10\ \rm{Gyr}$ time-scale.}
    \label{fig:orbitdirection}
\end{figure}

In Fig.~\ref{fig:period} we show the possible range of each dwarf galaxy's pericentre number by considering their orbits for all sampled current tangential velocities. The transparency of each bar represents the frequency of that particular number of pericentre passings over all sampled conditions for that dwarf galaxy. Those that are coloured in red (as well as Andromeda XII) are the 15 dwarf galaxies that lie in the great circle \citep{Ibata2013}. Dwarf galaxies whose current position places them far from the centre of M31 are only able to finish their first several cycles through a pericentre regardless of the tangential velocity we assume. Whereas other dwarf galaxies (e.g. Andromeda I, Andromeda IX) that are close to the M31 centre, are likely to pass through a pericentre far more frequently than those far away from the centre, given that they are in a low-energy orbit. These closer dwarf galaxies will hence explore much more of the phase space and mix their dynamic information. Furthermore, due to the asymmetric nature of the potential, the angular momentum of those dwarf galaxies will not be conserved. For the spherical potential, the asymmetric nature of the potential comes from the disc, hence the effect is more significant when a dwarf galaxy orbits close to the M31 centre. Naturally, this asymmetry is the reason that the prolate and oblate potentials shift the angular momentum even faster. For example, Fig.~\ref{fig:orbitdirection} shows variation of the angular momentum direction of a single Andromeda I orbit during $10\ \rm{Gyr}$ backwards integration throughout the spherical NFW profile. This particular dwarf galaxy will deviate from its plane of origin quickly due to the precession of its orbit. This effect dictates that it is unlikely that dwarf galaxies with high angular velocity and those with low angular velocity can remain contained within one stable asymmetric structure.

In consideration of this, we can determine that the 15 dwarf galaxies within the great plane are unlikely to be coherent or that this structure is much younger than $10\ \rm{Gyr}$ if some of its members have high angular velocity. A similar phenomenon happens to the asymmetric structure of the 23 dwarf galaxies that lie in the same hemisphere. Even though the dwarf galaxies far from M31 will stay in this hemisphere for a long time, dwarf galaxies close to M31 will leave it within a much shorter time-scale. One possible solution to satisfy the longevity of this structure over larger time-scales occurs if the dwarf galaxies close to M31 are in highly eccentric orbits with large tangential velocities. This would increase the orbit energy and decrease the angular velocities of those galaxies. As Fig.~\ref{fig:period} shows, the dwarf galaxies could always have low angular velocity as long as we assume that they have large enough tangential velocity.

The different positions of each dwarf galaxy within the potential require varying escape velocities. In our potential, both Andromeda XII and Andromeda XXVII do not exhibit any pericentre passings since they have very low escape velocities at their current position. \citet{Chapman2007} suggested that due to its large line-of-sight velocity relative to that of M31, Andromeda XII must be on its first infall into the system. Andromeda XXVII, is unbounded to the M31 in our model, which implies that it will never be a part of any long-term structure associated with the system. However, there is evidence that it is actually closer to us (and M31) \citep[e.g.][ and Preston et al. in prep]{Richardson2011,Conn2012} than the distance proposed in \citet{Conn2012}. This would allow its lowest possible energy to be lower so that it becomes bound to the M31. In this work though, these two dwarf galaxies will always be in long period orbits regardless of the tangential velocity we assign.

\begin{figure*}
    \includegraphics[width=\textwidth]{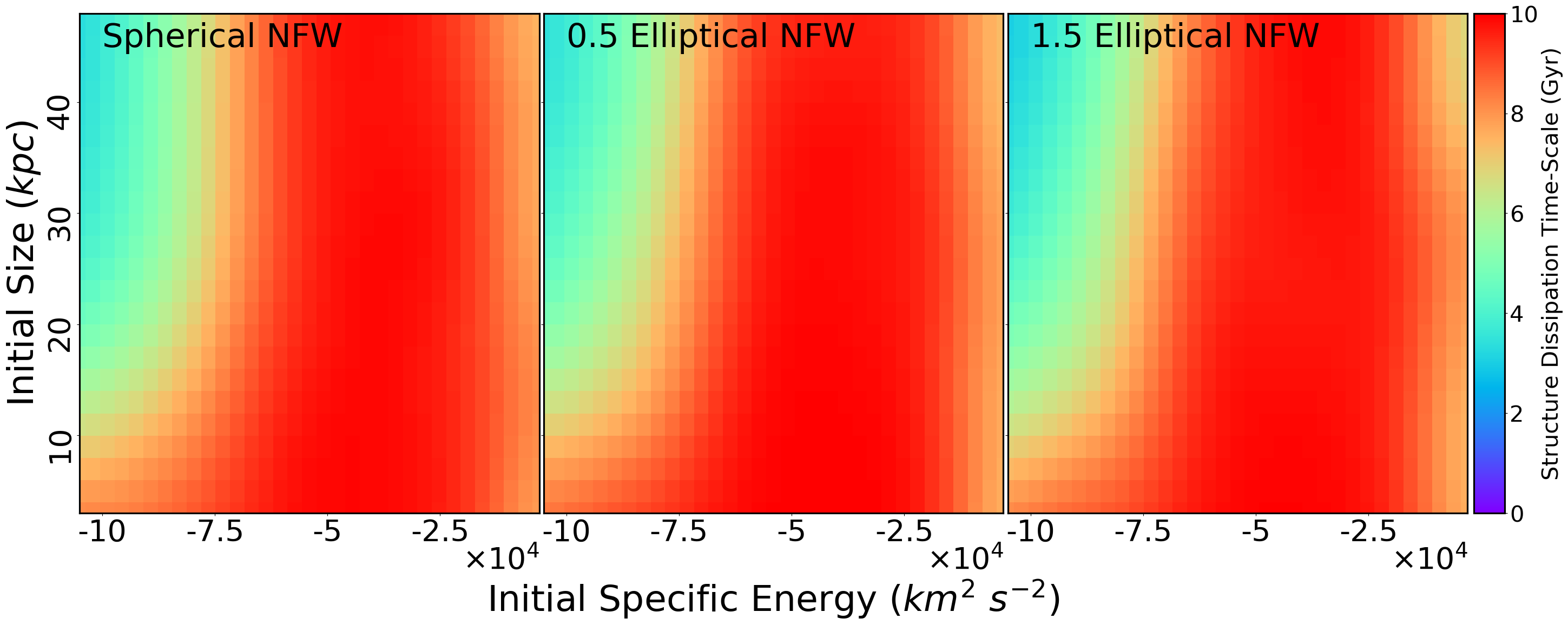}
    \vspace{-0.5cm}
    \caption{The time that associations with different initial sizes and specific energies need to dissipate to the extent that half of the dwarf galaxies within it are at least $90^{\circ}$ away from the association centre in angular distance. This plot is a Gaussian smoothed bin-map of our simulation results where each pixel spans $3200\ \rm{km^2\ s^{-2}}$ in specific energy and $2\ \rm{kpc}$ in size, with a Gaussian kernel $\sigma = 3\ \rm{pixels}$. A high-energy orbit is necessary for the association to be long-lived.}
    \label{fig:time_scale}
\end{figure*}

\subsection{Forward Integration of Orbits}
\label{sec:forward}

We attempt to construct the asymmetric distribution of dwarf galaxies around M31 by considering various associations of dwarf galaxies with different initial conditions and allow them to evolve for $10\ \rm{Gyr}$. For each association we place a dwarf galaxy at all 6 vertices of a regular octahedron and another dwarf galaxy at the centre of this group as a primary reference. We use the distance between the vertex dwarf galaxies and the central dwarf galaxy to represent the size of the association. The initial association sizes we take can be as large as $50\ \rm{kpc}$, with which we find the effect of a large association size is already clear. Then, under the condition that the central reference dwarf galaxy of each association is bound to the M31--MW system, we place these associations randomly within our potential and initialise them with a random velocity. Note that the 7 dwarf galaxies in each association (1 at the centre and 6 at the vertices) all possess the same initial velocity. We integrate each of the dwarf's orbits by setting the M31 halo potential to spherical, oblate, and prolate separately to investigate the effect of a non-spherical NFW profile as described in Sec.~\ref{sec:method}.

During each $10\ \rm{Gyr}$ integration, the associations will deform from their initial octahedral configuration to a more distorted shape. This change occurs due to the difference in potential energy across the association, since each dwarf galaxy within an association is initialised with the same velocity. The potential difference within each association is dependent upon the size of the association and the potential gradient surrounding it. The dispersion of the association is an effect of time-evolution that arises from the difference of this potential gradient. Based on the gravitational potential model (Eqs. \ref{PowerLaw}, \ref{MiyamotoNagai}, and \ref{NFW}) this difference is largest around the centre of M31. In the case where the association starts close to the centre of M31 with a low velocity, the dwarf galaxies in the association will be moving on low-energy orbits with a large gravitational potential and will complete a higher number of revolutions. Throughout a $10\ \rm{Gyr}$ integration of a dwarf galaxy association of this kind, the accumulative effect of the dispersion will be largely due to the considerable potential gradient in this region. Contrarily, an association that starts from a large radius and is moving on a high-energy orbit, will exhibit a smaller dispersive effect since the potential gradient is shallower along this orbit. 

\begin{figure*}
    \includegraphics[width=\textwidth]{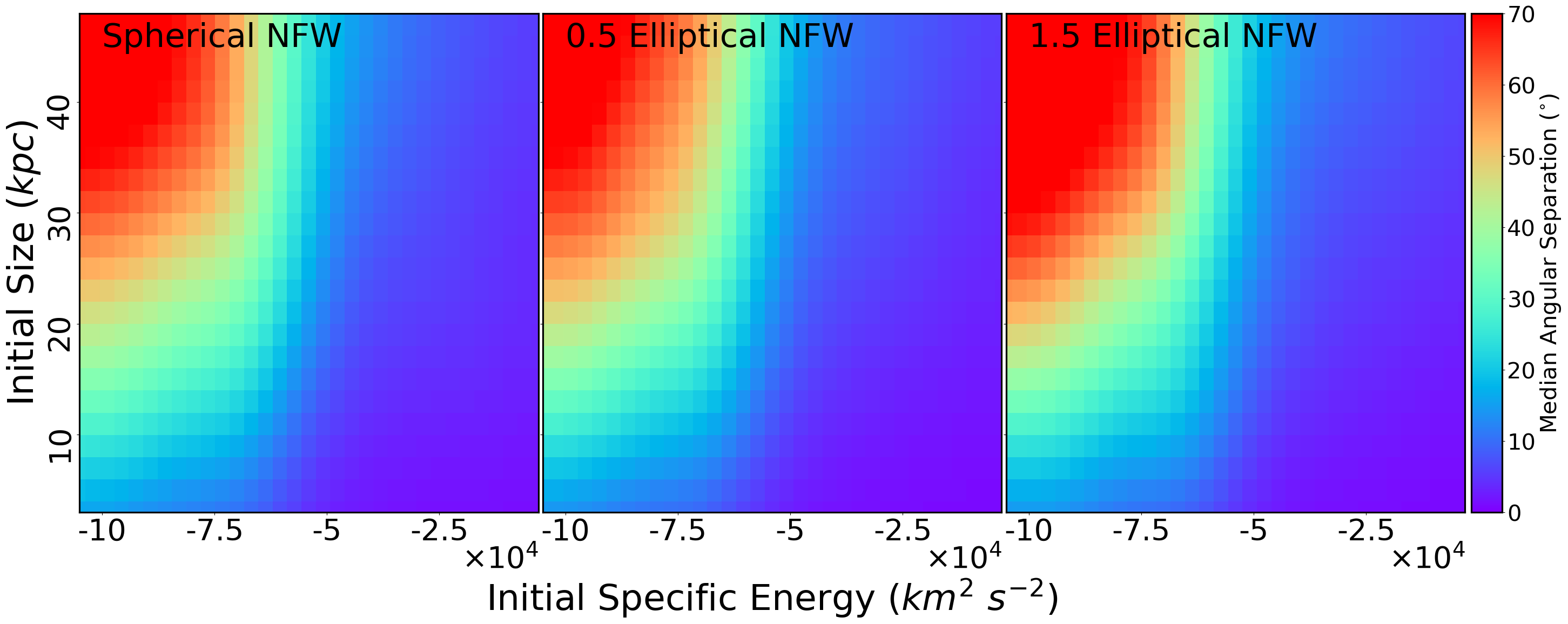}
    \includegraphics[width=\textwidth]{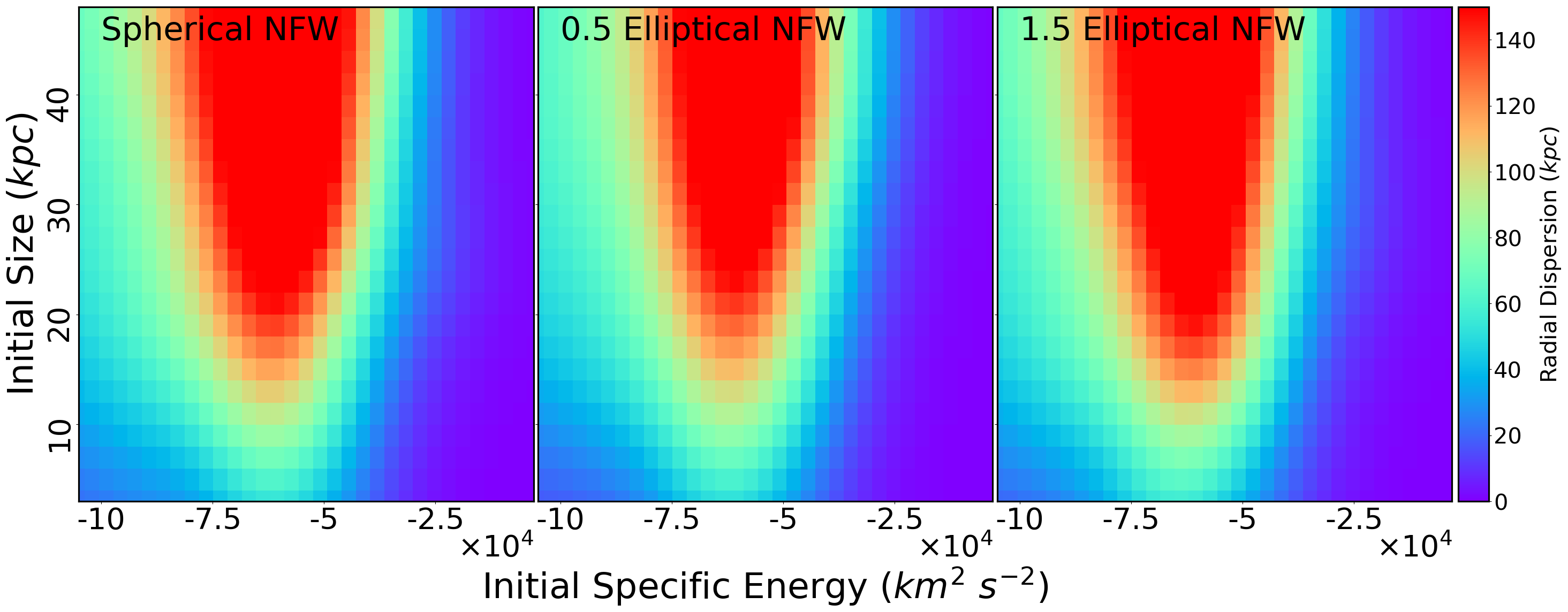}
    \includegraphics[width=\textwidth]{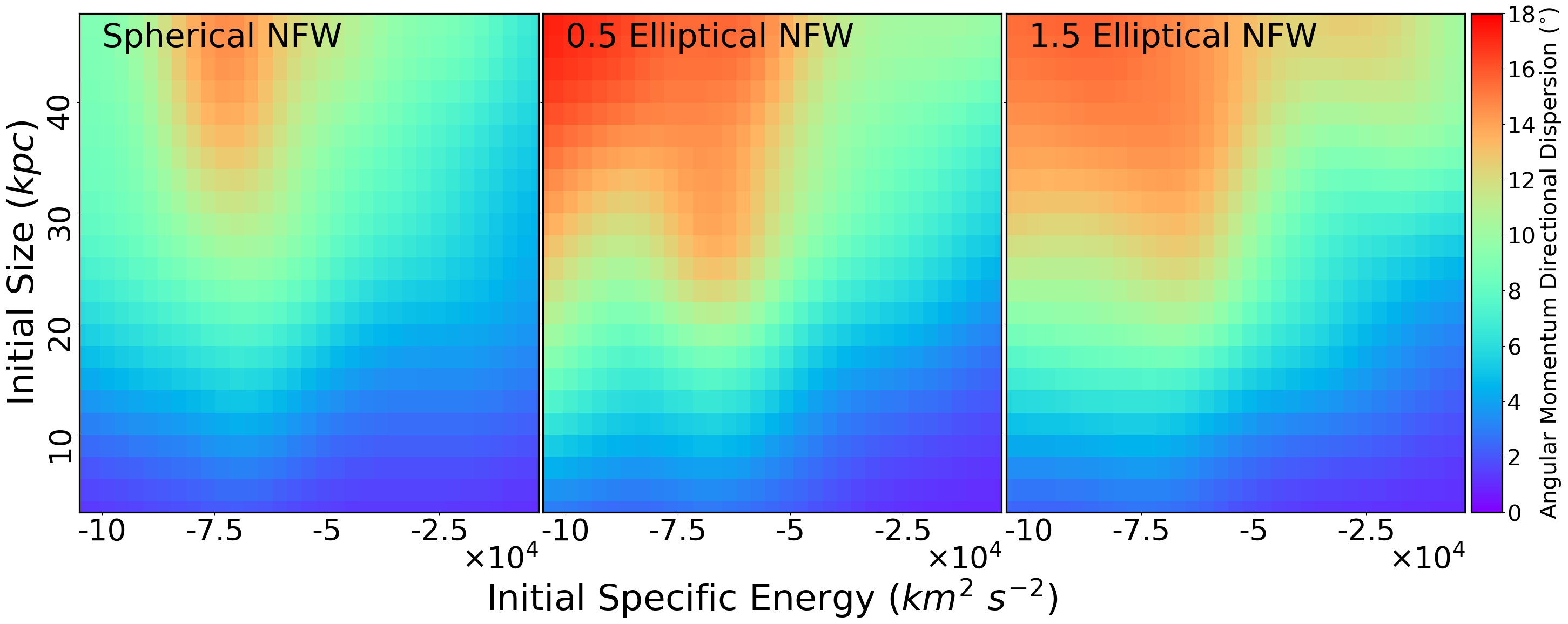}
    \vspace{-0.5cm}
    \caption{{\it Top}: The median angular distance of each dwarf galaxy to the centre of its association after $10\ \rm{Gyr}$ integration, sampled over associations varying in initial sizes and specific energies. {\it Middle}: The radial dispersion of associations after the $10\ \rm{Gyr}$ integration, sampled over associations varying in initial sizes and specific energies. A substantial radial dispersion occurs within an association after $10\ \rm{Gyr}$ if its initial size and energy are large. {\it Bottom}: The final angular momentum directional dispersion of all associations. For associations with initial sizes and energies that allow for the observed asymmetrical distributions (visualised in Figs.~\ref{fig:time_scale} and \ref{fig:dispersion}), the dispersion of the angular momentum direction is generally small. Note that we give the same velocity for each dwarf galaxy in an association, so this angular momentum dispersion is entirely dependent on the initial size of an association and the asymmetrical nature of the potential along its orbital path.}
    \label{fig:dispersion}
\end{figure*}

From these analyses we take the initial size and the total energy (the addition of the potential energy and the dynamic energy such that the total specific energy $\geq 0\ \rm{km^{2}/s^{2}}$ represents an unbounded orbit) as characteristics of the associations. We then consider the time-scale it takes for the association to dissipate to the extent that half of the galaxies in the association are at least $90^{\circ}$ away from the central reference in angular distance. The relationship between the size, energy, and dissipation time is presented in Fig.~\ref{fig:time_scale}. Associations that are moving on low-energy orbits with large sizes will be easily disrupted within shorter time-scales. The oblate and prolate NFW profiles have similar effects.
%
For each case, it is clear from this figure that for an association to remain compact after a $10\ \rm{Gyr}$ integration, it initially needs to be on a high-energy orbit with a preference towards having a small association size. The top panel of Fig.~\ref{fig:dispersion} similarly shows that to have most of the dwarf galaxies in one hemisphere after $10\ \rm{Gyr}$, as are the observed M31 dwarf galaxies, the association is required to be compact.

However, if the initial size is too small or the starting radius too large, the association will be unlikely to end up with a large difference in radius between its members. The middle panel of Fig.~\ref{fig:dispersion} shows the radial dispersion of the associations within the three potentials at the $10\ \rm{Gyr}$ snapshot given varying initial sizes and energies. For an association to obtain a large radial dispersion, it needs to initially exhibit both a high-energy and a large size that will not typically complete many revolutions of M31 within a $10\ \rm{Gyr}$ time-scale. We note that on the other hand, this energy cannot be too large otherwise most of the dwarf galaxies in the association will easily escape to $\geq 1\ \mathrm{Mpc}$ away from the M31 within $10\ \rm{Gyr}$.

\begin{figure*}
    \includegraphics[width=\textwidth]{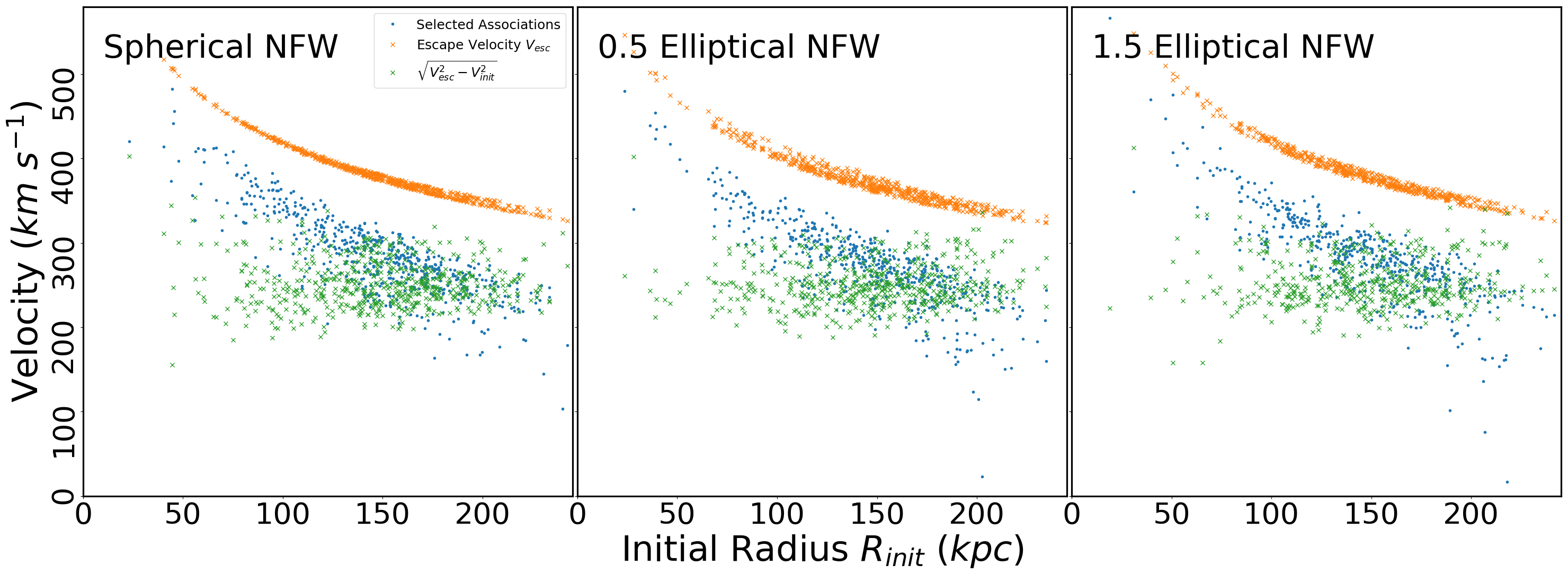}
    \vspace{-0.5cm}
    \caption{The initial radius ($R_{init}$) and initial velocity ($V_{init}$) of the selected associations that have final radial dispersion larger than $90\ \rm{kpc}$ and median angular separation smaller than $60^{\circ}$ (blue dots). The radial dispersion of association scales with the asymmetricity of the potential, and an association's angular separation is dependent on the steepness of the potential gradient. The orange crosses indicate the escape velocity ($V_{esc}$) of each association. The green crosses are the quantity $\sqrt{V_{esc}^{2} - V_{init}^2}$, which indicates the specific energy needed for the association to escape the potential well. This quantity also roughly indicates how far the association could reach from the M31 centre.}
    \label{fig:selected_sample}
\end{figure*}

We select associations that have final radial dispersion larger than $90\ \rm{kpc}$ and a final median angular separation smaller than $60^{\circ}$. The initial position and velocity distribution of these associations are shown in Fig.~\ref{fig:selected_sample}. We use orange crosses to indicate the escape velocity of each dwarf galaxy (indicated as $V_{esc}$) and green crosses to indicate $\sqrt{V_{esc}^{2} - V_{init}^{2}}$ of each association, a quantity which is the square root of the (negative) total specific energy of the association (note that for bounded orbits, this quantity is always positive). For all potentials, the selected association parameters typically occupy the high-energy region where the total specific energy $E = V_{esc}^{2} - V_{init}^{2}$ is roughly concentrated around $200^2 - 300^2\ \rm{km^2\ s^{-2}}$. ~\footnote{Note that the quantity $\frac{1}{2}(V_{esc}^{2} - V_{init}^{2})$ indicates the {\it negative} total specific energy}.

\begin{figure*}
    \includegraphics[width=\columnwidth]{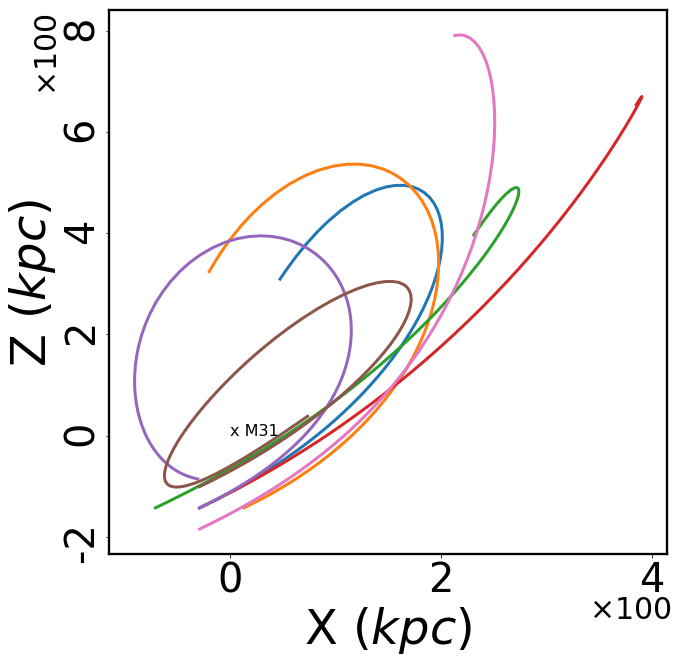}
    \includegraphics[width=\columnwidth]{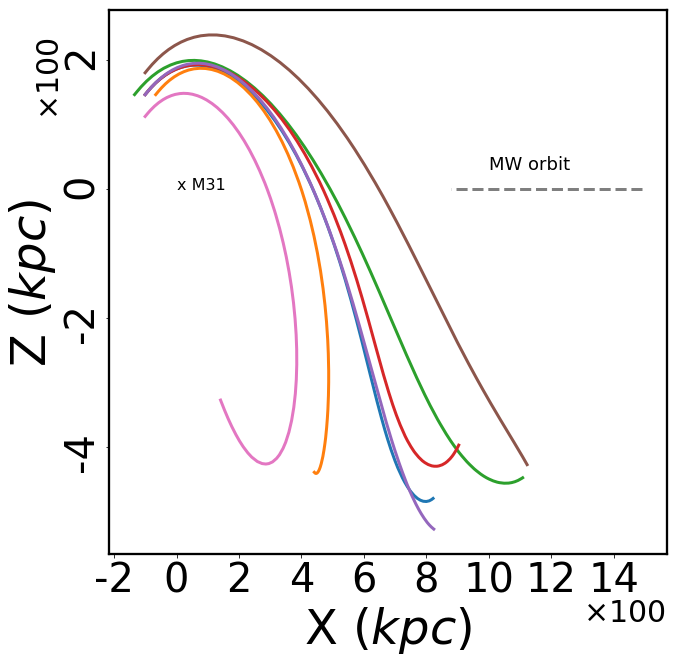}
    \caption{{\it Left}: Traced orbits of the members of an example association with an initial specific energy of $-28081\ \rm{km^2\ s^{-2}}$ and an initial size of $41.9\ \rm{kpc}$. The starting points of each dwarf galaxy are marked with crosses. This association is initially close to the M31 centre with a high velocity of $262.75\ \rm{km\ s^{-1}}$. Some members of the association acquire enough energy to escape to a large radius with highly eccentric orbits while others remain closely bound to M31. {\it Right}: Another example association with an initial specific energy of $-19381\ \rm{km^2\ s^{-2}}$ and an initial size of $33.8\ \rm{kpc}$. Some of the dwarf galaxies escape from M31 and fall towards the MW (the grey dashed line indicates the orbit of the MW) due to their high initial energy, which increases the angular momentum directional dispersion of these kinds of associations. }
    \label{fig:example_orbit}
\end{figure*}

The left panel of Fig.~\ref{fig:example_orbit} depicts the orbits of the members of one such association that has been placed $110\ \rm{kpc}$ away from the M31 centre with an initial velocity of $262.75\ \rm{km\ s^{-1}}$ and an initial size parameter of $41.9\ \rm{kpc}$. Some of the dwarf galaxies within this association are in a position of higher potential and after $10\ \rm{Gyr}$, 5 out of 7 dwarf galaxies in this association have acquired enough energy to escape to large radii and remain in the north hemisphere. The other 2 dwarf galaxies remain closely bound, within $100\ \rm{kpc}$ from the M31 centre. All of the dwarf galaxies in this association are in eccentric orbits. In fact, to construct the asymmetric distribution observed around M31 with a model that ignores dwarf-dwarf interaction such as this, a highly eccentric orbit is preferred so that some of the dwarf galaxies are close to the M31 centre while others are far away. Those associations with eccentric orbits either start close to the M31 centre with large initial velocities or far from the M31 centre with small velocities. Under both conditions, the associations will pass close to the M31 centre where the potential gradient is steep enough to radially separate the dwarf galaxies.

Since there is no dwarf-dwarf interaction in our model, the precession of the angular momentum of each dwarf galaxy is due to the asymmetric nature of the potentials. We show the final angular momentum directional dispersion of the associations in the bottom panel of Fig.~\ref{fig:dispersion}. Both the prolate and oblate NFW potentials lead to a significantly larger dispersion than when compared to the spherical potential.
The low-energy associations have larger angular velocities and revolve many times around the M31 during integration, so the accumulative effect is significant. For initial positions further from the M31 centre, and larger initial specific energies, the associations will have smaller angular velocities. Typically the potential field is weaker along the orbits of these dwarf galaxies, and hence the angular momentum directional dispersion is smaller than that of associations moving on low-energy orbits. Some of these high-energy associations with large initial sizes however, can acquire enough energy to escape the M31 potential and be captured by the MW. The right panel of the Fig.~\ref{fig:example_orbit} shows one of these kinds of associations, where some dwarf galaxies in this association are captured by the MW and change the direction of their velocity. This significantly changes the direction of their angular momentum. For an association to have a smaller final dispersion, a high initial energy and a small size are preferred because:

\begin{enumerate}
  \item We give the same initial velocity to each dwarf galaxy in an association. So a larger initial size will result in a larger initial angular momentum dispersion.
  \item For a spherical NFW profile, high-energy orbits will be far away from the centre where the asymmetricity of the disc component of the potential is small. We find that associations that result in an angular momentum distribution with a preferred direction are in high-energy orbits. However, if their energy is too high, some dwarf galaxies will be captured by the MW and significantly change the direction of their angular momentum.
\end{enumerate}

Similarly, we integrated the association orbits with the Hernquist potential \citep{Hernquist1990} as the form of the underlying dark halo, where the profile follows:
\begin{equation}
    \rho(r) = \frac{\rho_0}{(r/a)(1 + r/a)^3}
\end{equation}
This profile is scaled so that the total mass of the halo is $1\times 10^{12}\ \mathrm{M_{\odot}}$ and that the circular velocity at $(x,y,z) = (8,0,0)\ \mathrm{kpc}$ of the potential is $220\ \mathrm{km\ s^{-1}}$. The angular momentum directional dispersion are comparable to the spherical NFW profile. However, the resultant dissipation time-scale is much shorter, the median angular separation is more pronounced at higher specific energies, and the dwarf associations can conform to similar radial dispersions to the NFW profiles, though typically at larger energies too. This is expected since the Hernquist profile has a steeper gradient than compared to that of the spherical NFW potential. The initial size and specific energy parameter space that would allow for long-lasting coherent structures is extremely small if not non-existent and as such the Hernquist profile will destroy these structures within shorter-time scales.


\section{Discussion}
\label{sec:discussion}
The observed distribution of the 30 dwarf galaxies surrounding M31 is noticeably distinct, where 23 dwarf galaxies reside in one hemisphere and 15 dwarf galaxies are contained within a planar structure surrounding M31. The radial distribution of these dwarf galaxies, which ranges from $40$ to $400\ \rm{kpc}$, is less special. The asymmetric nature of this distribution raises the question of its origin. In previous sections, we explored the possible orbits of the dwarf galaxies around M31 in an attempt to examine the longevity of these structures as well as the possibility that these dwarf galaxies come from a single association. 

In our simulations, we include the gravitational potential of the MW which deforms the otherwise axisymmetric potential of M31 as in Fig.~\ref{fig:potential}, though its effect within $300\ \rm{kpc}$ is small. The integration results (e.g. Fig.~\ref{fig:example_orbit}) indicate that the MW potential will have some significant influence on the dwarf galaxy orbits from those high-energy associations with a large initial size.

The data we use for our analysis includes the sky position, distance, and line-of-sight velocity of each dwarf galaxy (as well as M31), giving us a 3-dimensional map and one component of the velocity of their distribution. Using this distribution as a current snapshot of the dwarf galaxies, we performed numerous integrations into the past for each dwarf galaxy's orbit by sampling over various tangential velocities. Through this method we find that there are only a limited number of possible bound orbits for those dwarf galaxies that are either far away from the M31 centre, or that have a high line-of-sight velocity. Lower energy orbits will be closer to the M31 centre with a higher angular velocity. From Fig.~\ref{fig:period} we see that some dwarf galaxies could possibly revolve the M31 centre in excess of 20 times within $10\ \rm{Gyr}$ corresponding to less than $500\ \rm{Myr}$ per revolution. In comparison, a few other dwarf galaxies could only complete up to 2 revolutions. Under the condition that these dwarf galaxies have come from a single large association where some members have been drawn into low-energy orbits and others into high-energy orbits, we find that the observed asymmetric structure will be short-lived (the life-time may be as short as $\sim 500\ \rm{Myr}$ during which some of the dwarf galaxies can complete one revolution of the M31). For this scenario to resemble the current snapshot of dwarf galaxies, the tangential velocity of those dwarf galaxies on lower energy orbits must be large. This is because there is little room to adjust the tangential velocities of those dwarf galaxies whose orbital energies have a high lower limit without making them unbound to the system. An increase in the magnitude of the tangential velocities of these low orbital energy dwarf galaxies is necessary to result in their orbits becoming highly eccentric with long periods. If the 23 dwarf galaxies that are in the same hemisphere are co-rotating around the M31, then their angular velocity (or the pericentre number in Fig.~\ref{fig:period}) should be roughly same as each other. By assuming all of the dwarf galaxies in the same hemisphere have the same specific energy as the Andromeda XXVIII (which has the largest specific energy without considering proper motion), we may calculate a rough estimation of the upper limit of the proper motion of the dwarf galaxies. As such, most of these dwarf galaxies will have a tangential velocity of $\sim 150 - 350\ \mathrm{km\ s^{-1}}$ which corresponds to a magnitude of proper motion $\sim 45 - 110\ \mathrm{\mu as\ yr^{-1}}$, which agrees with a recent proper motion estimation based on the planer structure\citep{Hodkinson2019}. Under these conditions Andromeda XVII, which is currently the closest dwarf to M31, would need to have a tangential velocity of $\sim 382\ \mathrm{km\ s^{-1}}$. Note that these estimations only take into account the dwarfs' motion relative to the M31 and as such, the actual proper motion we observe would differ from this and would include the effects of both the magnitude and direction of the proper motion of M31 \citep[e.g.][]{vanderMarel2019}. These values are the contribution of the proper motion from the dwarfs motion with respect to the M31. Given the magnitude of these proper motions, this may be detectable within the next generation of telescopes.

Having established that the observed asymmetries are likely short-lived, we explored whether they could have formed from the recent infall of a single association. To do this, we sampled associations with different initial conditions, placed them in the M31--MW potential, and integrated their orbits forwards for $10\ \rm{Gyr}$. Here we use the initial size and specific energy to characterise each association. We find that, for an association to result in the observed small angular dispersion after integration, it requires a high initial energy, and that the observed large radial dispersion requires a high initial energy and a large initial size. We also find that the orbital energies of the associations that result in the same hemisphere structure are concentrated where the total specific energy $E = V_{esc}^{2} - V_{init}^{2} \sim 200^2 - 300^2\  \rm{km^2\ s^{-2}}$. This energy is high enough that the dwarf galaxies could reach further than $500\ \rm{kpc}$ from the centre, but still be bound to M31. Compared to lower energies, this high-energy keeps the angular momentum directional dispersion small so that the disc structure can also survive. Because we give the same velocity to all dwarf galaxies in an association, the initial angular momentum differences are due to the size of the association as well as it's position within our potential model. Then the precession of the dwarf galaxy orbits originates from the asymmetricity of the potential as well as the  comparative initial size of the association relative to its initial distance from the M31 centre -- which magnifies the differential apsidal precession within the group. Far from the centre of M31 (with the spherical NFW potential), the potential is approximately spherically symmetric and the initial size of the association contributes less to the differential apsidal precession. As such, the precession of those high-energy orbits is relatively small. However, both the prolate and oblate NFW potentials will significantly destroy the planar structure by changing the direction of the angular momentum of the dwarf galaxies with their asymmetric shape.

In Fig.~\ref{fig:example_orbit} we show one example association orbit that shows angular asymmetry. Here, 5 of the 7 dwarf galaxies of that association end up in the northern hemisphere and some dwarf galaxies remain within $200\ \rm{kpc}$ from the M31 centre while others are far away. The high-energy initial condition is necessary from two aspects. Firstly, some of the dwarf galaxies in this association need to end up far from the M31 centre by the end of the integration; and secondly, the angular velocity of the association cannot be so large that the dwarf galaxies become well mixed. From both the forward and backward integrations, we find that for a long-living asymmetric structure to exist, the association likely needs to have recently completed or currently be in its first revolution of M31.

In reality, it seems unlikely that a structure with a size of $\sim 40\ \rm{kpc}$ has formed $110\ \rm{kpc}$ away from the M31 centre with a velocity of $\sim 260\ \rm{km\ s^{-1}}$. One possible case is that such an association has formed earlier and subsequently fallen inward to the M31 centre from large radius. In this scenario, the structure of the association would become disturbed by the M31 potential and may be capable of developing into the initial condition of the association modelled in Fig.~\ref{fig:example_orbit}. The infalling orbit would need to be eccentric to a large enough degree that the association could approach close to the M31 centre. This way the stronger tidal forces would cause some dwarf galaxies to become more bound to M31. Additionally, the size of this association could not be too small to ensure that enough dwarf galaxies are still able to escape to large radii and resemble the currently observed large radial dispersion. In conclusion, the intersection of initial condition regions required by the observed radial dispersion, angular asymmetry and planar structure is small. The asymmetric structures -- especially the planar structure, which will be easily destroyed by both the prolate and oblate potentials -- are less likely to have come from a single association than not, or they are short-lived.

We note that these results and subsequent discussion are based on the assumption that this asymmetric structure has originated from a single association that has been able to last for up to $10\ \rm{Gyr}$. The initial conditions may be utterly distinct from these assumptions if this is a young structure, or if this asymmetric structure is merely a coincidence. Another premise of the simulation results presented here is that the dwarf galaxies are non-interacting throughout the course of their orbits. Our results demonstrate how the M31--MW potential alone could contribute to the observed asymmetric distribution under the assumptions previously discussed. However, the interaction between each dwarf galaxy may establish more self-bounded associations than those that appear in our analysis. This interaction could also provide a means of transferring energy and angular momentum between the dwarf galaxies of an association so that a large radial dispersion can evolve. The interaction may also provide a mechanism for a longer-lasting distributional asymmetricity of the dwarfs. An association with sufficient self-gravity could be held together during its first infall so that it may exhibit a condition similar to the initial conditions that we have shown could result in the observed distribution of dwarf galaxies. A possible candidate for the progenitor of such an association is NGC3190 \citep{Bellazzini2013} as it is large and currently far from M31. It is possible that the observed distribution of dwarf galaxies could have resulted from an association centred on a younger NGC3190 that had passed close enough to M31 to be tidally disrupted. We leave this concept and the inclusion of dwarf-dwarf interaction in our model for future work. In addition, the M33 could have a significant contribution to the potential as well, similar to the effect of Large Magellanic Cloud on the MW. A more detailed model of this system with M33 could also be investigated in the future and would complement the research we present here. We leave other dynamical effects such as the time-dependency of the potential components and the incorporation of an appropriate cosmological setting for future work as well. Thus, the research we present here, whilst based on simple assumptions, is a map for future works from which we may compare the effects of these various factors.

\section*{Acknowledgements}
ZW gratefully acknowledges financial support through the Dean's International Postgraduate Research Scholarship from the Physics School of the University of Sydney.
GFL acknowledges support from the 
Institute of Advanced Studies Santander Fellowship at the University of Surrey and thanks them for their hospitality where the initial idea for this project was conceived.



\bibliographystyle{mnras}
\bibliography{M31}

\begin{thebibliography}{}
\makeatletter
\relax
\def\mn@urlcharsother{\let\do\@makeother \do\$\do\&\do\#\do\^\do\_\do\%\do\~}
\def\mn@doi{\begingroup\mn@urlcharsother \@ifnextchar [ {\mn@doi@}
  {\mn@doi@[]}}
\def\mn@doi@[#1]#2{\def\@tempa{#1}\ifx\@tempa\@empty \href
  {http://dx.doi.org/#2} {doi:#2}\else \href {http://dx.doi.org/#2} {#1}\fi
  \endgroup}
\def\mn@eprint#1#2{\mn@eprint@#1:#2::\@nil}
\def\mn@eprint@arXiv#1{\href {http://arxiv.org/abs/#1} {{\tt arXiv:#1}}}
\def\mn@eprint@dblp#1{\href {http://dblp.uni-trier.de/rec/bibtex/#1.xml}
  {dblp:#1}}
\def\mn@eprint@#1:#2:#3:#4\@nil{\def\@tempa {#1}\def\@tempb {#2}\def\@tempc
  {#3}\ifx \@tempc \@empty \let \@tempc \@tempb \let \@tempb \@tempa \fi \ifx
  \@tempb \@empty \def\@tempb {arXiv}\fi \@ifundefined
  {mn@eprint@\@tempb}{\@tempb:\@tempc}{\expandafter \expandafter \csname
  mn@eprint@\@tempb\endcsname \expandafter{\@tempc}}}

\bibitem[\protect\citeauthoryear{{Bahl} \& {Baumgardt}}{{Bahl} \&
  {Baumgardt}}{2014}]{Bahl2014}
{Bahl} H.,  {Baumgardt} H.,  2014, \mn@doi [\mnras] {10.1093/mnras/stt2399},
  \href {https://ui.adsabs.harvard.edu/abs/2014MNRAS.438.2916B} {438, 2916}

\bibitem[\protect\citeauthoryear{{Bellazzini}, {Oosterloo}, {Fraternali}  \&
  {Beccari}}{{Bellazzini} et~al.}{2013}]{Bellazzini2013}
{Bellazzini} M.,  {Oosterloo} T.,  {Fraternali} F.,   {Beccari} G.,  2013,
  \mn@doi [\aap] {10.1051/0004-6361/201322744}, \href
  {https://ui.adsabs.harvard.edu/abs/2013A&A...559L..11B} {559, L11}

\bibitem[\protect\citeauthoryear{{Binney} \& {Tremaine}}{{Binney} \&
  {Tremaine}}{2008}]{Binney2008}
{Binney} J.,  {Tremaine} S.,  2008, {Galactic Dynamics: Second Edition}.
Princeton University Press

\bibitem[\protect\citeauthoryear{{Bovy}}{{Bovy}}{2015}]{Bovy2015}
{Bovy} J.,  2015, \mn@doi [\apjs] {10.1088/0067-0049/216/2/29}, \href
  {http://adsabs.harvard.edu/abs/2015ApJS..216...29B} {216, 29}

\bibitem[\protect\citeauthoryear{{Bovy} \& {Rix}}{{Bovy} \&
  {Rix}}{2013}]{Bovy2013}
{Bovy} J.,  {Rix} H.-W.,  2013, \mn@doi [\apj] {10.1088/0004-637X/779/2/115},
  \href {http://adsabs.harvard.edu/abs/2013ApJ...779..115B} {779, 115}

\bibitem[\protect\citeauthoryear{{Bovy} et~al.,}{{Bovy}
  et~al.}{2012}]{Bovy2012}
{Bovy} J.,  et~al., 2012, \mn@doi [\apj] {10.1088/0004-637X/759/2/131}, \href
  {http://adsabs.harvard.edu/abs/2012ApJ...759..131B} {759, 131}

\bibitem[\protect\citeauthoryear{{Bowden}, {Evans}  \& {Belokurov}}{{Bowden}
  et~al.}{2013}]{Bowden2013}
{Bowden} A.,  {Evans} N.~W.,   {Belokurov} V.,  2013, \mn@doi [\mnras]
  {10.1093/mnras/stt1253}, \href
  {http://adsabs.harvard.edu/abs/2013MNRAS.435..928B} {435, 928}

\bibitem[\protect\citeauthoryear{{Bowden}, {Evans}  \& {Belokurov}}{{Bowden}
  et~al.}{2014}]{Bowden2014}
{Bowden} A.,  {Evans} N.~W.,   {Belokurov} V.,  2014, \mn@doi [\apj]
  {10.1088/2041-8205/793/2/L42}, \href
  {https://ui.adsabs.harvard.edu/abs/2014ApJ...793L..42B} {793, L42}

\bibitem[\protect\citeauthoryear{{Buck}, {Macci{\`o}}  \& {Dutton}}{{Buck}
  et~al.}{2015}]{Buck2015}
{Buck} T.,  {Macci{\`o}} A.~V.,   {Dutton} A.~A.,  2015, \mn@doi [\apj]
  {10.1088/0004-637X/809/1/49}, \href
  {http://adsabs.harvard.edu/abs/2015ApJ...809...49B} {809, 49}

\bibitem[\protect\citeauthoryear{Buck, Dutton  \& Macciò}{Buck
  et~al.}{2016}]{Buck2016}
Buck T.,  Dutton A.~A.,   Macciò A.~V.,  2016, \mn@doi [Monthly Notices of the
  Royal Astronomical Society] {10.1093/mnras/stw1232}, 460, 4348

\bibitem[\protect\citeauthoryear{Cautun, Wang, Frenk  \& Sawala}{Cautun
  et~al.}{2015a}]{Cautun2015a}
Cautun M.,  Wang W.,  Frenk C.~S.,   Sawala T.,  2015a, \mn@doi [Monthly
  Notices of the Royal Astronomical Society] {10.1093/mnras/stv490}, 449, 2576

\bibitem[\protect\citeauthoryear{{Cautun}, {Bose}, {Frenk}, {Guo}, {Han},
  {Hellwing}, {Sawala}  \& {Wang}}{{Cautun} et~al.}{2015b}]{Cautun2015b}
{Cautun} M.,  {Bose} S.,  {Frenk} C.~S.,  {Guo} Q.,  {Han} J.,  {Hellwing}
  W.~A.,  {Sawala} T.,   {Wang} W.,  2015b, \mn@doi [\mnras]
  {10.1093/mnras/stv1557}, \href
  {https://ui.adsabs.harvard.edu/abs/2015MNRAS.452.3838C} {452, 3838}

\bibitem[\protect\citeauthoryear{{Chapman} et~al.,}{{Chapman}
  et~al.}{2007}]{Chapman2007}
{Chapman} S.~C.,  et~al., 2007, \mn@doi [\apj] {10.1086/519377}, \href
  {https://ui.adsabs.harvard.edu/abs/2007ApJ...662L..79C} {662, L79}

\bibitem[\protect\citeauthoryear{{Clemens}}{{Clemens}}{1985}]{Clemens1985}
{Clemens} D.~P.,  1985, \mn@doi [\apj] {10.1086/163386}, \href
  {https://ui.adsabs.harvard.edu/abs/1985ApJ...295..422C} {295, 422}

\bibitem[\protect\citeauthoryear{{Conn} et~al.,}{{Conn}
  et~al.}{2012}]{Conn2012}
{Conn} A.~R.,  et~al., 2012, \mn@doi [\apj] {10.1088/0004-637X/758/1/11}, \href
  {https://ui.adsabs.harvard.edu/abs/2012ApJ...758...11C} {758, 11}

\bibitem[\protect\citeauthoryear{{Conn} et~al.,}{{Conn}
  et~al.}{2013}]{Conn2013}
{Conn} A.~R.,  et~al., 2013, \mn@doi [\apj] {10.1088/0004-637X/766/2/120},
  \href {https://ui.adsabs.harvard.edu/abs/2013ApJ...766..120C} {766, 120}

\bibitem[\protect\citeauthoryear{{D'Onghia} \& {Lake}}{{D'Onghia} \&
  {Lake}}{2008}]{D'Onghia2008}
{D'Onghia} E.,  {Lake} G.,  2008, \mn@doi [\apjl] {10.1086/592995}, \href
  {https://ui.adsabs.harvard.edu/abs/2008ApJ...686L..61D} {686, L61}

\bibitem[\protect\citeauthoryear{{Deason} et~al.,}{{Deason}
  et~al.}{2012}]{Deason2012}
{Deason} A.~J.,  et~al., 2012, \mn@doi [\mnras]
  {10.1111/j.1365-2966.2012.21639.x}, \href
  {https://ui.adsabs.harvard.edu/abs/2012MNRAS.425.2840D} {425, 2840}

\bibitem[\protect\citeauthoryear{{Debattista}, {Moore}, {Quinn}, {Kazantzidis},
  {Maas}, {Mayer}, {Read}  \& {Stadel}}{{Debattista}
  et~al.}{2008}]{Debattista2008}
{Debattista} V.~P.,  {Moore} B.,  {Quinn} T.,  {Kazantzidis} S.,  {Maas} R.,
  {Mayer} L.,  {Read} J.,   {Stadel} J.,  2008, \mn@doi [\apj]
  {10.1086/587977}, \href
  {https://ui.adsabs.harvard.edu/abs/2008ApJ...681.1076D} {681, 1076}

\bibitem[\protect\citeauthoryear{{Dehnen} \& {Binney}}{{Dehnen} \&
  {Binney}}{1998}]{Dehnen1998}
{Dehnen} W.,  {Binney} J.,  1998, \mn@doi [\mnras]
  {10.1046/j.1365-8711.1998.01282.x}, \href
  {http://adsabs.harvard.edu/abs/1998MNRAS.294..429D} {294, 429}

\bibitem[\protect\citeauthoryear{{Dubinski}}{{Dubinski}}{1994}]{Dubinski1994}
{Dubinski} J.,  1994, \mn@doi [\apj] {10.1086/174512}, \href
  {https://ui.adsabs.harvard.edu/abs/1994ApJ...431..617D} {431, 617}

\bibitem[\protect\citeauthoryear{{Forero-Romero} \& {Arias}}{{Forero-Romero} \&
  {Arias}}{2018}]{Forero-Romero2018}
{Forero-Romero} J.~E.,  {Arias} V.,  2018, \mn@doi [\mnras]
  {10.1093/mnras/sty1349}, \href
  {http://adsabs.harvard.edu/abs/2018MNRAS.478.5533F} {478, 5533}

\bibitem[\protect\citeauthoryear{{Gaia Collaboration} et~al.,}{{Gaia
  Collaboration} et~al.}{2018}]{GaiaCollaboration2018}
{Gaia Collaboration} et~al., 2018, \mn@doi [\aap]
  {10.1051/0004-6361/201832698}, \href
  {https://ui.adsabs.harvard.edu/abs/2018A&A...616A..12G} {616, A12}

\bibitem[\protect\citeauthoryear{Gillet, Ocvirk, Aubert, Knebe, Libeskind,
  Yepes, Gottl{\"{o}}ber  \& Hoffman}{Gillet et~al.}{2015}]{Gillet2015}
Gillet N.,  Ocvirk P.,  Aubert D.,  Knebe A.,  Libeskind N.,  Yepes G.,
  Gottl{\"{o}}ber S.,   Hoffman Y.,  2015, \mn@doi [\apj]
  {10.1088/0004-637X/800/1/34}, 800

\bibitem[\protect\citeauthoryear{{Goerdt}, {Burkert}  \& {Ceverino}}{{Goerdt}
  et~al.}{2013}]{Goerdt2013}
{Goerdt} T.,  {Burkert} A.,   {Ceverino} D.,  2013, arXiv e-prints, \href
  {https://ui.adsabs.harvard.edu/abs/2013arXiv1307.2102G} {p. arXiv:1307.2102}

\bibitem[\protect\citeauthoryear{Gong et~al.,}{Gong et~al.}{2019}]{Gong2019}
Gong C.~C.,  et~al., 2019, \mn@doi [\mnras] {10.1093/mnras/stz1917}

\bibitem[\protect\citeauthoryear{{Hayashi} \& {Chiba}}{{Hayashi} \&
  {Chiba}}{2014}]{Hayashi2014}
{Hayashi} K.,  {Chiba} M.,  2014, \mn@doi [\apj] {10.1088/0004-637X/789/1/62},
  \href {https://ui.adsabs.harvard.edu/abs/2014ApJ...789...62H} {789, 62}

\bibitem[\protect\citeauthoryear{{Hernquist}}{{Hernquist}}{1990}]{Hernquist1990}
{Hernquist} L.,  1990, \mn@doi [\apj] {10.1086/168845}, \href
  {https://ui.adsabs.harvard.edu/abs/1990ApJ...356..359H} {356, 359}

\bibitem[\protect\citeauthoryear{{Hodkinson} \& {Scholtz}}{{Hodkinson} \&
  {Scholtz}}{2019}]{Hodkinson2019}
{Hodkinson} B.,  {Scholtz} J.,  2019, \mn@doi [\mnras] {10.1093/mnras/stz1893},
  \href {https://ui.adsabs.harvard.edu/abs/2019MNRAS.488.3231H} {488, 3231}

\bibitem[\protect\citeauthoryear{{Holmberg} \& {Flynn}}{{Holmberg} \&
  {Flynn}}{2000}]{Holmberg2000}
{Holmberg} J.,  {Flynn} C.,  2000, \mn@doi [\mnras]
  {10.1046/j.1365-8711.2000.02905.x}, \href
  {https://ui.adsabs.harvard.edu/abs/2000MNRAS.313..209H} {313, 209}

\bibitem[\protect\citeauthoryear{{Ibata}, {Lewis}, {Totten}  \&
  {Irwin}}{{Ibata} et~al.}{1998}]{Ibata1998}
{Ibata} R.~A.,  {Lewis} G.~F.,  {Totten} E.,   {Irwin} M.~J.,  1998, in
  {Kroupa} P.,  {Palous} J.,   {Spurzem} R.,  eds, Dynamical Studies of Star
  Clusters and Galaxies. p.~178

\bibitem[\protect\citeauthoryear{{Ibata} et~al.,}{{Ibata}
  et~al.}{2013}]{Ibata2013}
{Ibata} R.~A.,  et~al., 2013, \mn@doi [\nat] {10.1038/nature11717}, \href
  {http://adsabs.harvard.edu/abs/2013Natur.493...62I} {493, 62}

\bibitem[\protect\citeauthoryear{{Ibata}, {Ibata}, {Lewis}, {Martin}, {Conn},
  {Elahi}, {Arias}  \& {Fernando}}{{Ibata} et~al.}{2014}]{Ibata2014}
{Ibata} R.~A.,  {Ibata} N.~G.,  {Lewis} G.~F.,  {Martin} N.~F.,  {Conn} A.,
  {Elahi} P.,  {Arias} V.,   {Fernando} N.,  2014, \mn@doi [\apj]
  {10.1088/2041-8205/784/1/L6}, \href
  {https://ui.adsabs.harvard.edu/abs/2014ApJ...784L...6I} {784, L6}

\bibitem[\protect\citeauthoryear{Jones, Oliphant, Peterson  et~al.}{Jones
  et~al.}{2001}]{Eric2001}
Jones E.,  Oliphant T.,  Peterson P.,   et~al., 2001, {SciPy}: Open source
  scientific tools for {Python}, \url {http://www.scipy.org/}

\bibitem[\protect\citeauthoryear{{Kafle}, {Sharma}, {Lewis}  \& {Bland
  -Hawthorn}}{{Kafle} et~al.}{2012}]{Kafle2012}
{Kafle} P.~R.,  {Sharma} S.,  {Lewis} G.~F.,   {Bland -Hawthorn} J.,  2012,
  \mn@doi [\apj] {10.1088/0004-637X/761/2/98}, \href
  {https://ui.adsabs.harvard.edu/abs/2012ApJ...761...98K} {761, 98}

\bibitem[\protect\citeauthoryear{{Kafle}, {Sharma}, {Lewis}  \&
  {Bland-Hawthorn}}{{Kafle} et~al.}{2014}]{Kafle2014}
{Kafle} P.~R.,  {Sharma} S.,  {Lewis} G.~F.,   {Bland-Hawthorn} J.,  2014,
  \mn@doi [\apj] {10.1088/0004-637X/794/1/59}, \href
  {https://ui.adsabs.harvard.edu/abs/2014ApJ...794...59K} {794, 59}

\bibitem[\protect\citeauthoryear{{Kroupa}, {Theis}  \& {Boily}}{{Kroupa}
  et~al.}{2005}]{Kroupa2005}
{Kroupa} P.,  {Theis} C.,   {Boily} C.~M.,  2005, \mn@doi [\aap]
  {10.1051/0004-6361:20041122}, \href
  {http://adsabs.harvard.edu/abs/2005A%26A...431..517K} {431, 517}

\bibitem[\protect\citeauthoryear{{Li} \& {Helmi}}{{Li} \&
  {Helmi}}{2009}]{Li2009}
{Li} Y.-S.,  {Helmi} A.,  2009, in {Andersen} J.,  {Nordstr{\"o}ara} {m} B.,
  {Bland-Hawthorn} J.,  eds,  IAU Symposium Vol. 254, The Galaxy Disk in
  Cosmological Context. pp 263--268 (\mn@eprint {arXiv} {0807.2780}),
  \mn@doi{10.1017/S1743921308027683}

\bibitem[\protect\citeauthoryear{{Li} \& {White}}{{Li} \&
  {White}}{2008}]{Li2008}
{Li} Y.-S.,  {White} S. D.~M.,  2008, \mn@doi [\mnras]
  {10.1111/j.1365-2966.2007.12748.x}, \href
  {https://ui.adsabs.harvard.edu/abs/2008MNRAS.384.1459L} {384, 1459}

\bibitem[\protect\citeauthoryear{Libeskind, Knebe, Hoffman, Gottl{\"{o}}ber,
  Yepes  \& Steinmetz}{Libeskind et~al.}{2011}]{Libeskind2011}
Libeskind N.~I.,  Knebe A.,  Hoffman Y.,  Gottl{\"{o}}ber S.,  Yepes G.,
  Steinmetz M.,  2011, \mn@doi [\mnras] {10.1111/j.1365-2966.2010.17786.x},
  411, 1525

\bibitem[\protect\citeauthoryear{Libeskind, Hoffman, Tully, Courtois,
  Pomar{\`{e}}de, Gottl{\"{o}}ber  \& Steinmetz}{Libeskind
  et~al.}{2015}]{Libeskind2015}
Libeskind N.~I.,  Hoffman Y.,  Tully R.~B.,  Courtois H.~M.,  Pomar{\`{e}}de
  D.,  Gottl{\"{o}}ber S.,   Steinmetz M.,  2015, \mn@doi [\mnras]
  {10.1093/mnras/stv1302}, 452, 1052

\bibitem[\protect\citeauthoryear{{Libeskind}, {Guo}, {Tempel}  \&
  {Ibata}}{{Libeskind} et~al.}{2016}]{Libeskind2016}
{Libeskind} N.~I.,  {Guo} Q.,  {Tempel} E.,   {Ibata} R.,  2016, \mn@doi [\apj]
  {10.3847/0004-637X/830/2/121}, \href
  {https://ui.adsabs.harvard.edu/abs/2016ApJ...830..121L} {830, 121}

\bibitem[\protect\citeauthoryear{{Lovell}, {Eke}, {Frenk}  \&
  {Jenkins}}{{Lovell} et~al.}{2011}]{Lovell2011}
{Lovell} M.~R.,  {Eke} V.~R.,  {Frenk} C.~S.,   {Jenkins} A.,  2011, \mn@doi
  [\mnras] {10.1111/j.1365-2966.2011.18377.x}, \href
  {https://ui.adsabs.harvard.edu/abs/2011MNRAS.413.3013L} {413, 3013}

\bibitem[\protect\citeauthoryear{{Lux}, {Read}  \& {Lake}}{{Lux}
  et~al.}{2010}]{Lux2010}
{Lux} H.,  {Read} J.~I.,   {Lake} G.,  2010, in {Debattista} V.~P.,  {Popescu}
  C.~C.,  eds,  American Institute of Physics Conference Series Vol. 1240,
  American Institute of Physics Conference Series. pp 415--416 (\mn@eprint
  {arXiv} {0912.3276}), \mn@doi{10.1063/1.3458551}

\bibitem[\protect\citeauthoryear{{Lynden-Bell}}{{Lynden-Bell}}{1976}]{Lynden-Bell1976}
{Lynden-Bell} D.,  1976, \mn@doi [\mnras] {10.1093/mnras/174.3.695}, \href
  {http://adsabs.harvard.edu/abs/1976MNRAS.174..695L} {174, 695}

\bibitem[\protect\citeauthoryear{{Lynden-Bell} \& {Lynden-Bell}}{{Lynden-Bell}
  \& {Lynden-Bell}}{1995}]{Lynden-Bell1995}
{Lynden-Bell} D.,  {Lynden-Bell} R.~M.,  1995, \mn@doi [\mnras]
  {10.1093/mnras/275.2.429}, \href
  {http://adsabs.harvard.edu/abs/1995MNRAS.275..429L} {275, 429}

\bibitem[\protect\citeauthoryear{Maji, Zhu, Marinacci  \& Li}{Maji
  et~al.}{2017}]{Maji2017}
Maji M.,  Zhu Q.,  Marinacci F.,   Li Y.,  2017, \mn@doi [The Astrophysical
  Journal] {10.3847/1538-4357/aa72f5}, 843, 62

\bibitem[\protect\citeauthoryear{{McClure-Griffiths} \&
  {Dickey}}{{McClure-Griffiths} \& {Dickey}}{2007}]{McClure-Griffiths2007}
{McClure-Griffiths} N.~M.,  {Dickey} J.~M.,  2007, \mn@doi [\apj]
  {10.1086/522297}, \href
  {https://ui.adsabs.harvard.edu/abs/2007ApJ...671..427M} {671, 427}

\bibitem[\protect\citeauthoryear{{McConnachie}}{{McConnachie}}{2012}]{McConnachie2012}
{McConnachie} A.~W.,  2012, \mn@doi [\aj] {10.1088/0004-6256/144/1/4}, \href
  {http://adsabs.harvard.edu/abs/2012AJ....144....4M} {144, 4}

\bibitem[\protect\citeauthoryear{{McConnachie} \& {Irwin}}{{McConnachie} \&
  {Irwin}}{2006}]{McConnachie2006}
{McConnachie} A.~W.,  {Irwin} M.~J.,  2006, \mn@doi [\mnras]
  {10.1111/j.1365-2966.2005.09771.x}, \href
  {https://ui.adsabs.harvard.edu/abs/2006MNRAS.365..902M} {365, 902}

\bibitem[\protect\citeauthoryear{{McConnachie} et~al.,}{{McConnachie}
  et~al.}{2009}]{McConnachie2009}
{McConnachie} A.~W.,  et~al., 2009, \mn@doi [\nat] {10.1038/nature08327}, \href
  {http://adsabs.harvard.edu/abs/2009Natur.461...66M} {461, 66}

\bibitem[\protect\citeauthoryear{{Metz}, {Kroupa}  \& {Jerjen}}{{Metz}
  et~al.}{2007}]{Metz2007}
{Metz} M.,  {Kroupa} P.,   {Jerjen} H.,  2007, \mn@doi [\mnras]
  {10.1111/j.1365-2966.2006.11228.x}, \href
  {http://adsabs.harvard.edu/abs/2007MNRAS.374.1125M} {374, 1125}

\bibitem[\protect\citeauthoryear{{Metz}, {Kroupa}  \& {Libeskind}}{{Metz}
  et~al.}{2008}]{Metz2008}
{Metz} M.,  {Kroupa} P.,   {Libeskind} N.~I.,  2008, \mn@doi [\apj]
  {10.1086/587833}, \href
  {https://ui.adsabs.harvard.edu/abs/2008ApJ...680..287M} {680, 287}

\bibitem[\protect\citeauthoryear{{Navarro}, {Frenk}  \& {White}}{{Navarro}
  et~al.}{1996}]{Navarro1996}
{Navarro} J.~F.,  {Frenk} C.~S.,   {White} S. D.~M.,  1996, \mn@doi [\apj]
  {10.1086/177173}, \href
  {https://ui.adsabs.harvard.edu/abs/1996ApJ...462..563N} {462, 563}

\bibitem[\protect\citeauthoryear{{Pasetto} \& {Chiosi}}{{Pasetto} \&
  {Chiosi}}{2009}]{Pasetto2009}
{Pasetto} S.,  {Chiosi} C.,  2009, \mn@doi [\aap]
  {10.1051/0004-6361/200811153}, \href
  {http://adsabs.harvard.edu/abs/2009A%26A...499..385P} {499, 385}

\bibitem[\protect\citeauthoryear{{Pawlowski} \& {Kroupa}}{{Pawlowski} \&
  {Kroupa}}{2013}]{Pawlowski2013}
{Pawlowski} M.~S.,  {Kroupa} P.,  2013, \mn@doi [\mnras]
  {10.1093/mnras/stt1429}, \href
  {http://adsabs.harvard.edu/abs/2013MNRAS.435.2116P} {435, 2116}

\bibitem[\protect\citeauthoryear{{Pawlowski}, {Pflamm-Altenburg}  \&
  {Kroupa}}{{Pawlowski} et~al.}{2012}]{Pawlowski2012}
{Pawlowski} M.~S.,  {Pflamm-Altenburg} J.,   {Kroupa} P.,  2012, \mn@doi
  [\mnras] {10.1111/j.1365-2966.2012.20937.x}, \href
  {http://adsabs.harvard.edu/abs/2012MNRAS.423.1109P} {423, 1109}

\bibitem[\protect\citeauthoryear{{Pawlowski}, {Bullock}, {Kelley}  \&
  {Famaey}}{{Pawlowski} et~al.}{2019}]{Pawlowski2019}
{Pawlowski} M.~S.,  {Bullock} J.~S.,  {Kelley} T.,   {Famaey} B.,  2019,
  \mn@doi [\apj] {10.3847/1538-4357/ab10e0}, \href
  {https://ui.adsabs.harvard.edu/abs/2019ApJ...875..105P} {875, 105}

\bibitem[\protect\citeauthoryear{Penarrubia, Gómez, Besla, Erkal  \&
  Ma}{Penarrubia et~al.}{2015}]{Penarrubia2015}
Penarrubia J.,  Gómez F.~A.,  Besla G.,  Erkal D.,   Ma Y.-Z.,  2015, \mn@doi
  [Monthly Notices of the Royal Astronomical Society: Letters]
  {10.1093/mnrasl/slv160}, 456, L54

\bibitem[\protect\citeauthoryear{Phillips, Cooper, Bullock  \&
  Boylan-Kolchin}{Phillips et~al.}{2015}]{Phillips2015}
Phillips J.~I.,  Cooper M.~C.,  Bullock J.~S.,   Boylan-Kolchin M.,  2015,
  \mn@doi [Monthly Notices of the Royal Astronomical Society]
  {10.1093/mnras/stv1770}, 453, 3839

\bibitem[\protect\citeauthoryear{Posti \& Helmi}{Posti \&
  Helmi}{2019}]{Posti2019}
Posti L.,  Helmi A.,  2019, A\&A, 621

\bibitem[\protect\citeauthoryear{{Read}}{{Read}}{2014}]{Read2014}
{Read} J.~I.,  2014, \mn@doi [Journal of Physics G Nuclear Physics]
  {10.1088/0954-3899/41/6/063101}, \href
  {https://ui.adsabs.harvard.edu/abs/2014JPhG...41f3101R} {41, 063101}

\bibitem[\protect\citeauthoryear{{Read}, {Lake}, {Agertz}  \&
  {Debattista}}{{Read} et~al.}{2008}]{Read2008}
{Read} J.~I.,  {Lake} G.,  {Agertz} O.,   {Debattista} V.~P.,  2008, \mn@doi
  [\mnras] {10.1111/j.1365-2966.2008.13643.x}, \href
  {https://ui.adsabs.harvard.edu/abs/2008MNRAS.389.1041R} {389, 1041}

\bibitem[\protect\citeauthoryear{{Richardson} et~al.,}{{Richardson}
  et~al.}{2011}]{Richardson2011}
{Richardson} J.~C.,  et~al., 2011, \mn@doi [\apj] {10.1088/0004-637X/732/2/76},
  \href {https://ui.adsabs.harvard.edu/abs/2011ApJ...732...76R} {732, 76}

\bibitem[\protect\citeauthoryear{{Sch{\"o}nrich}, {Binney}  \&
  {Dehnen}}{{Sch{\"o}nrich} et~al.}{2010}]{Schonrich2010}
{Sch{\"o}nrich} R.,  {Binney} J.,   {Dehnen} W.,  2010, \mn@doi [\mnras]
  {10.1111/j.1365-2966.2010.16253.x}, \href
  {https://ui.adsabs.harvard.edu/abs/2010MNRAS.403.1829S} {403, 1829}

\bibitem[\protect\citeauthoryear{{Seigar}, {Barth}  \& {Bullock}}{{Seigar}
  et~al.}{2008}]{Seigar2008}
{Seigar} M.~S.,  {Barth} A.~J.,   {Bullock} J.~S.,  2008, \mn@doi [\mnras]
  {10.1111/j.1365-2966.2008.13732.x}, \href
  {https://ui.adsabs.harvard.edu/abs/2008MNRAS.389.1911S} {389, 1911}

\bibitem[\protect\citeauthoryear{{Shao}, {Cautun}, {Frenk}, {Gao}, {Crain},
  {Schaller}, {Schaye}  \& {Theuns}}{{Shao} et~al.}{2016}]{Shao2016}
{Shao} S.,  {Cautun} M.,  {Frenk} C.~S.,  {Gao} L.,  {Crain} R.~A.,  {Schaller}
  M.,  {Schaye} J.,   {Theuns} T.,  2016, \mn@doi [\mnras]
  {10.1093/mnras/stw1247}, \href
  {https://ui.adsabs.harvard.edu/abs/2016MNRAS.460.3772S} {460, 3772}

\bibitem[\protect\citeauthoryear{Simon}{Simon}{2018}]{Simon2018}
Simon J.~D.,  2018, \mn@doi [The Astrophysical Journal]
  {10.3847/1538-4357/aacdfb}, 863, 89

\bibitem[\protect\citeauthoryear{Sohn, Watkins, Fardal, van~der Marel, Deason,
  Besla  \& Bellini}{Sohn et~al.}{2018}]{Sohn2018}
Sohn S.~T.,  Watkins L.~L.,  Fardal M.~A.,  van~der Marel R.~P.,  Deason A.~J.,
   Besla G.,   Bellini A.,  2018, \mn@doi [The Astrophysical Journal]
  {10.3847/1538-4357/aacd0b}, 862, 52

\bibitem[\protect\citeauthoryear{{Wang}, {Frenk}  \& {Cooper}}{{Wang}
  et~al.}{2013}]{Wang2013}
{Wang} J.,  {Frenk} C.~S.,   {Cooper} A.~P.,  2013, \mn@doi [\mnras]
  {10.1093/mnras/sts442}, \href
  {https://ui.adsabs.harvard.edu/abs/2013MNRAS.429.1502W} {429, 1502}

\bibitem[\protect\citeauthoryear{{Wang}, {Guo}, {Libeskind}, {Tempel}, {Wei}
  \& {Kang}}{{Wang} et~al.}{2019}]{Wang2019}
{Wang} P.,  {Guo} Q.,  {Libeskind} N.~I.,  {Tempel} E.,  {Wei} C.,   {Kang} X.,
   2019, \mn@doi [\mnras] {10.1093/mnras/stz285}, \href
  {https://ui.adsabs.harvard.edu/abs/2019MNRAS.484.4325W} {484, 4325}

\bibitem[\protect\citeauthoryear{Watkins, Evans  \& An}{Watkins
  et~al.}{2010}]{Watkins2010}
Watkins L.~L.,  Evans N.~W.,   An J.~H.,  2010, \mn@doi [Monthly Notices of the
  Royal Astronomical Society] {10.1111/j.1365-2966.2010.16708.x}, 406, 264

\bibitem[\protect\citeauthoryear{{Wegg}, {Gerhard}  \& {Bieth}}{{Wegg}
  et~al.}{2019}]{Wegg2019}
{Wegg} C.,  {Gerhard} O.,   {Bieth} M.,  2019, \mn@doi [\mnras]
  {10.1093/mnras/stz572}, \href
  {https://ui.adsabs.harvard.edu/abs/2019MNRAS.485.3296W} {485, 3296}

\bibitem[\protect\citeauthoryear{{Xue} et~al.,}{{Xue} et~al.}{2008}]{Xue2008}
{Xue} X.~X.,  et~al., 2008, \mn@doi [\apj] {10.1086/589500}, \href
  {https://ui.adsabs.harvard.edu/abs/2008ApJ...684.1143X} {684, 1143}

\bibitem[\protect\citeauthoryear{{Zentner}, {Kravtsov}, {Gnedin}  \&
  {Klypin}}{{Zentner} et~al.}{2005}]{Zentner2005}
{Zentner} A.~R.,  {Kravtsov} A.~V.,  {Gnedin} O.~Y.,   {Klypin} A.~A.,  2005,
  \mn@doi [\apj] {10.1086/431355}, \href
  {https://ui.adsabs.harvard.edu/abs/2005ApJ...629..219Z} {629, 219}

\bibitem[\protect\citeauthoryear{{Zhang}, {Rix}, {van de Ven}, {Bovy}, {Liu}
  \& {Zhao}}{{Zhang} et~al.}{2013}]{Zhang2013}
{Zhang} L.,  {Rix} H.-W.,  {van de Ven} G.,  {Bovy} J.,  {Liu} C.,   {Zhao} G.,
   2013, \mn@doi [\apj] {10.1088/0004-637X/772/2/108}, \href
  {http://adsabs.harvard.edu/abs/2013ApJ...772..108Z} {772, 108}

\bibitem[\protect\citeauthoryear{{de Vaucouleurs}}{{de
  Vaucouleurs}}{1958}]{deVaucouleurs1958}
{de Vaucouleurs} G.,  1958, \mn@doi [\apj] {10.1086/146564}, \href
  {http://adsabs.harvard.edu/abs/1958ApJ...128..465D} {128, 465}

\bibitem[\protect\citeauthoryear{van~der Marel, Fardal, Sohn, Patel, Besla, del
  Pino, Sahlmann  \& Watkins}{van~der Marel et~al.}{2019}]{vanderMarel2019}
van~der Marel R.~P.,  Fardal M.~A.,  Sohn S.~T.,  Patel E.,  Besla G.,  del
  Pino A.,  Sahlmann J.,   Watkins L.~L.,  2019, \mn@doi [The Astrophysical
  Journal] {10.3847/1538-4357/ab001b}, 872, 24

\makeatother
\end{thebibliography}



\newpage
\appendix

\section{More figures}

\begin{figure*}
    \includegraphics[width=2\columnwidth]{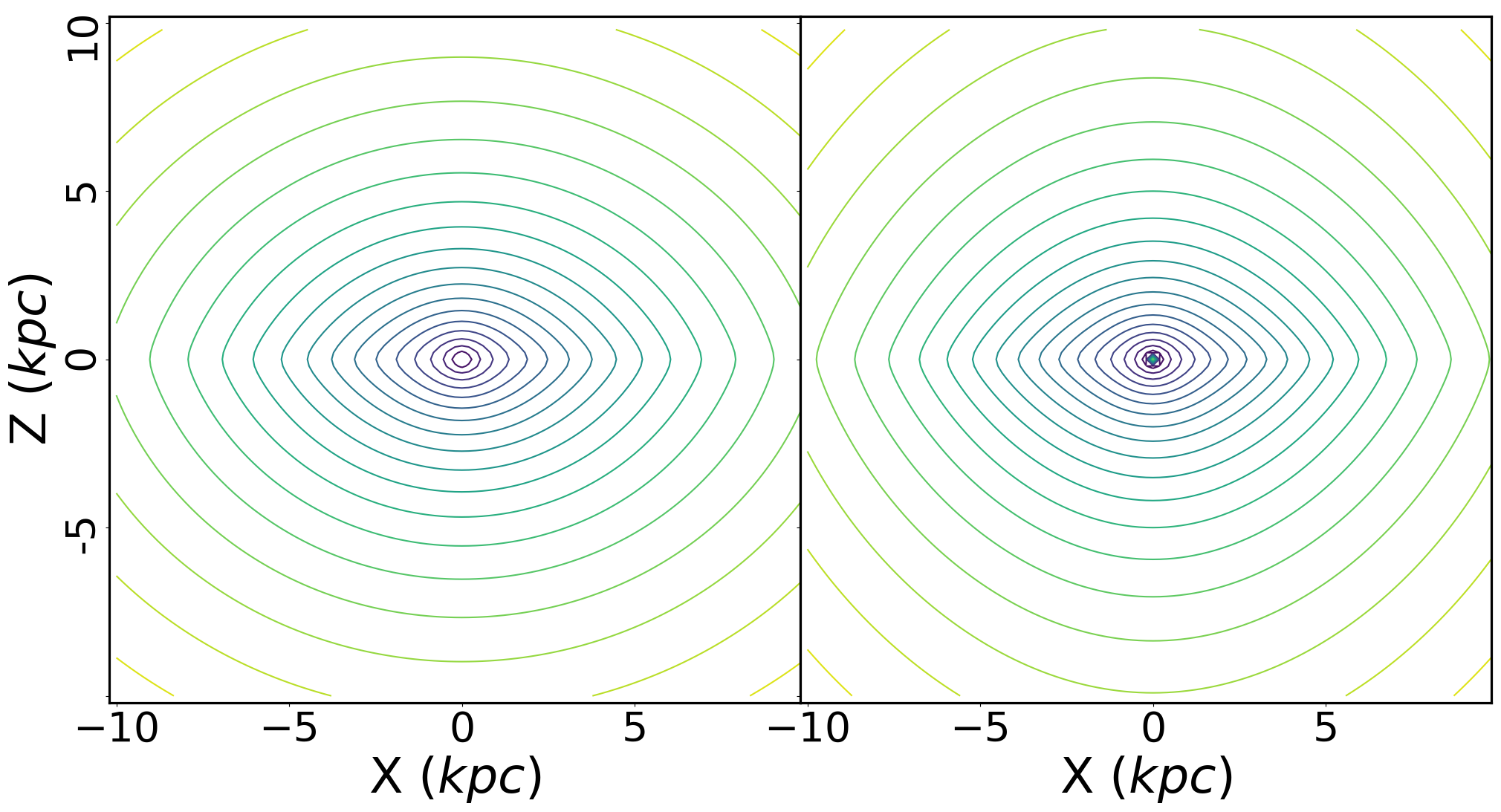}
        \caption{Equi-potential contours for the oblate NFW profile ({\it Left}), and the prolate NFW profile ({\it Right}).}
        \label{fig:potential_Triaxial}
\end{figure*}

\begin{figure*}
    \centering
    \includegraphics[width=2\columnwidth]{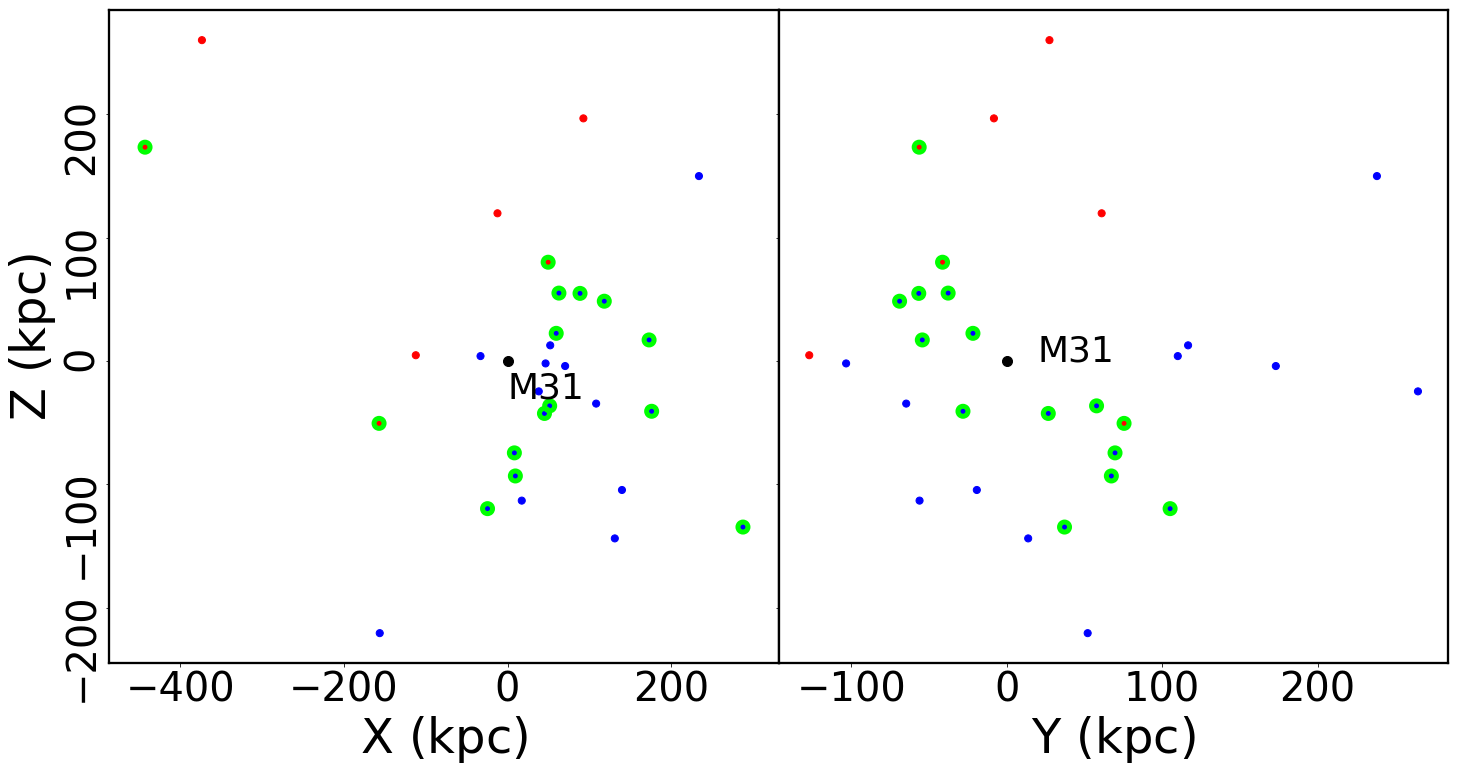}
    \caption{The Cartesian projection of the dwarf galaxies in the M31. The points centred on the lime circles are the 15 dwarf galaxies contained within the thin disk, and the blue dots are the 23 dwarf galaxies that are within the same hemisphere.}
    \label{fig:3Dprojection}
\end{figure*}


\bsp	
\label{lastpage}
\end{document}